\documentstyle[12pt]{article}

\font\twlgot =eufm10 scaled \magstep1
\font\egtgot =eufm8
\font\sevgot =eufm7
\font\twlmsb =msbm10 scaled \magstep1
\font\egtmsb =msbm8
\font\sevmsb =msbm7

\newfam\gotfam
\def\pgot{\fam\gotfam\twlgot}
\textfont\gotfam\twlgot
\scriptfont\gotfam\egtgot
\scriptscriptfont\gotfam\sevgot
\def\got{\protect\pgot}
\newfam\msbfam
\textfont\msbfam\twlmsb
\scriptfont\msbfam\egtmsb
\scriptscriptfont\msbfam\sevmsb
\def\Bbb{\protect\pBbb}
\def\pBbb{\relax\ifmmode\expandafter\Bb\else\typeout{You cann't use
Bbb in text mode}\fi}
\def\Bb #1{{\fam\msbfam\relax#1}}

\newcommand{\gO}{{\got O}}
\newcommand{\gQ}{{\got T}}

\newcommand{\gE}{{\got E}}
\newcommand{\gA}{{\got A}}

\newcommand{\gd}{{\got d}}

\newcommand{\gS}{{\got S}}

\def\thebibliography#1{\section*{References}\list
  {[\arabic{enumi}]}{\settowidth\labelwidth{#1}\leftmargin\labelwidth
    \advance\leftmargin\labelsep
    \usecounter{enumi}}
    \def\newblock{\hskip .11em plus .33em minus .07em}
    \sloppy\clubpenalty4000\widowpenalty4000
    \sfcode`\.=1000\relax}

\def\op#1{\mathop{\fam0 #1}\limits}

\newcommand{\Ker}{{\rm Ker\,}}
\newcommand{\im}{{\rm Im\,}}
\newcommand{\nm}[1]{|{#1}|}
\newcommand{\beq}{\begin{equation}}
\newcommand{\eeq}{\end{equation}}
\newcommand{\ben}{\begin{eqnarray}}
\newcommand{\een}{\end{eqnarray}}
\newcommand{\be}{\begin{eqnarray*}}
\newcommand{\ee}{\end{eqnarray*}}
\newcommand{\bea}{\begin{eqalph}}
\newcommand{\eea}{\end{eqalph}}
\newcommand{\cA}{{\cal A}}

\newcommand{\cP}{{\cal P}}

\newcommand{\cL}{{\cal L}}

\newcommand{\cV}{{\cal V}}

\newcommand{\cS}{{\cal S}}
\newcommand{\cC}{{\cal C}}
\newcommand{\cO}{{\cal O}}

\newcommand{\bL}{{\bf L}}

\newcommand{\bs}{{\bf s}}

\newcommand{\al}{\alpha}
\newcommand{\vr}{\varrho}
\newcommand{\bt}{\beta}
\newcommand{\dl}{\delta}
\newcommand{\la}{\lambda}
\newcommand{\La}{\Lambda}
\newcommand{\f}{\phi}
\newcommand{\om}{\omega}

\newcommand{\m}{\mu}

\newcommand{\g}{\gamma}
\newcommand{\G}{\Gamma}
\newcommand{\th}{\theta}

\newcommand{\vt}{\vartheta}
\newcommand{\vf}{\varphi}
\newcommand{\up}{\upsilon}
\newcommand{\lng}{\langle}
\newcommand{\rng}{\rangle}
\newcommand{\di}{{\rm dim\,}}

\newcommand{\si}{\sigma}
\newcommand{\Si}{\Sigma}
\newcommand{\w}{\wedge}
\newcommand{\wt}{\widetilde}

\newcommand{\ol}{\overline}
\newcommand{\dr}{\partial}
\newcommand{\ar}{\op\longrightarrow}
\newcommand{\ot}{\otimes}
\newcommand{\ap}{\approx}

\let\ssection=\section
\renewcommand{\section}{\setcounter{equation}{0}\ssection}

\newcounter{eqalph}[section]
\newcounter{equationa}[section]
\newcounter{example}[section]
\newcounter{remark}[section]
\newcounter{theorem}[section]
\newcounter{proposition}[section]
\newcounter{lemma}[section]
\newcounter{corollary}[section]
\newcounter{definition}[section]

\def\theremark{\arabic{section}.\arabic{remark}}

\def\thedefinition{\arabic{section}.\arabic{definition}}

\newenvironment{proof}{\noindent 
{\it Proof.}}{$\Box$ \medskip}
\newenvironment{rem}{\refstepcounter{remark}\medskip\noindent{\it
Remark \theremark.}}{\medskip}
\newenvironment{theo}{\refstepcounter{definition} 
\bigskip\noindent{\bf Theorem \thedefinition.} \it}{\medskip}
\newenvironment{prop}{\refstepcounter{definition} 
\bigskip\noindent{\bf Proposition \thedefinition.}\it}{\medskip}
\newenvironment{lem}{\refstepcounter{definition} 
\bigskip\noindent{\bf Lemma \thedefinition.}\it}{\medskip}
\newenvironment{cor}{\refstepcounter{definition} 
\bigskip\noindent{\bf Corollary \thedefinition.}\it}{\medskip}

\newenvironment{eqalph}{\stepcounter{equation}
\setcounter{equationa}{\value{equation}}
\setcounter{equation}{0}

\begin{eqnarray}}{\end{eqnarray}
\setcounter{equation}{\value{equationa}}}

\newcommand{\mar}[1]{}

\begin{document}
\hbox{}

{\parindent=0pt

{\large \bf Lagrangian 
supersymmetries depending on derivatives. Global analysis and
cohomology}
\bigskip

{\bf Giovanni Giachetta}$^1$, {\bf Luigi 
Mangiarotti}$^1$, {\bf Gennadi Sardanashvily}$^2$
\bigskip

\begin{small}

$^1$ Department of Mathematics and Informatics, University
of Camerino, 62032 Camerino (MC), Italy

\medskip

$^2$ Department of Theoretical Physics, Physics Faculty,
Moscow State University, 117234 Moscow, Russia
\bigskip

\end{small}

{\bf Abstract:} 
Lagrangian contact supersymmetries (depending on
derivatives of arbitrary order) are treated in very general
setting. The cohomology of the  variational
bicomplex on an arbitrary graded manifold and
the iterated cohomology of a generic nilpotent contact
supersymmetry are computed. In particular, the first
variational formula and conservation laws for Lagrangian
systems on graded manifolds using contact supersymmetries are
obtained.
  }

\section{Introduction}

At present, BRST transformations in the BV formalism
\cite{bat,gom} provide the most interesting example of
Lagrangian contact supersymmetries, depending on derivatives
and preserving the contact ideal of graded exterior forms.
Much that is already known regarding Lagrangian BRST
theory (including the short variational complex, BRST
cohomology \cite{barn95,barn,bran}, Noether's conservation
laws \cite{barn,fat,fulp})  has been formulated in terms of jet
manifolds of vector bundles (see
\cite{barn}  for a survey) since the jet manifold formalism
provides the algebraic description of Lagrangian and
Hamiltonian systems of both even and odd variables. In spite
of this formulation, most authors however  assume the base
manifold
$X$ of these bundles to be contractible because, e.g., the
relative (local in the terminology of
\cite{barn95,barn}) cohomology are not trivial even when $X=\Bbb
R^n$.  

Stimulated by the BRST theory,  
we consider Lagrangian systems
of odd variables and contact supersymmetries in very general
setting. For this purpose, one usually calls
into play fiber bundles over 
supermanifolds 
\cite{cari,cia,frank,mont}. We describe odd variables and their
jets on an arbitrary smooth manifold
$X$ as generating elements of the structure ring of a
graded manifold whose body is
$X$ \cite{book00,ijmp,mpla}. This definition 
differs from that of jets of a graded
fiber bundle \cite{hern}, but 
reproduces  the heuristic notion of jets of ghosts
in the field-antifield BRST theory on $\Bbb R^n$
\cite{barn,bran01}. 

Our goal is the following. Firstly, we construct the
$\Bbb Z_2$-graded variational bicomplex  on a graded manifold
with an arbitrary body $X$, and obtain the 
cohomology of its short variational subcomplex and the
complex of one-contact graded forms (Theorem
\ref{g96}). In particular, the first
variational formula and conservation laws for Lagrangian
systems on graded manifolds using contact supersymmetries are
obtained (formulae (\ref{g107}) -- (\ref{g108})). 

Secondly, the
iterated cohomology of a generic nilpotent contact
supersymmetry is computed (Theorems
\ref{g250}, \ref{04100} and \ref{g251}). In the most
interesting case of the form degree $n=\di X$, it coincides
with the above mentioned relative cohomology. Therefore, 
we extend the results of \cite{barn} and our recent work
\cite{lmp} to an arbitrary nilpotent contact
supersymmetry.

As is well-known, generalized (depending on derivatives)
symmetries of differential equations have been intensively
investigated
\cite{and93,bry,ibr,kras,olv}.   
Generalized symmetries of Lagrangian systems on a
local coordinate domain 
have been described in detail
\cite{bry,olv}. 
The variational bicomplex constructed in 
the framework 
of the infinite order jet formalism
enables one to provide the 
global analysis of 
Lagrangian systems on a fiber bundle and their symmetries 
\cite{ander,jmp,book00,tak2}. Sketched in Section
2 of our work, this analysis is extended to Lagrangian systems
on graded manifolds (Section 3).

Recall that an
$r$-order Lagrangian on a fiber bundle $Y\to
X$ is defined as a
horizontal density
$L:J^rY\to
\op\w^nT^*X$,
$n=\di X$, on the $r$-order jet manifold $J^rY$ of sections
of $Y\to X$.
 With 
the inverse system of 
finite order jet manifolds
\mar{5.10}\beq
X\op\longleftarrow^\pi Y\op\longleftarrow^{\pi^1_0} J^1Y
\longleftarrow \cdots J^{r-1}Y \op\longleftarrow^{\pi^r_{r-1}}
J^rY\longleftarrow\cdots,
\label{5.10}
\eeq
 we have
the direct system
\mar{5.7}\beq
\cO^*(X)\op\longrightarrow^{\pi^*} \cO^*(Y) 
\op\longrightarrow^{\pi^1_0{}^*} \cO_1^* \ar\cdots \cO^*_{r-1}
\op\longrightarrow^{\pi^r_{r-1}{}^*}
 \cO_r^* \longrightarrow\cdots \label{5.7}
\eeq
of graded differential algebras (henceforth GDAs) of exterior
forms on jet manifolds with respect to 
the pull-back monomorphisms $\pi^r_{r-1}{}^*$. Its direct
limit
is the GDA $\cO_\infty^*$
consisting of all the exterior forms on
finite order jet manifolds modulo the pull-back identification.
The exterior differential on $\cO_\infty^*$ is decomposed into
the sum $d=d_H+d_V$ of the total and the vertical 
differentials. These differentials split $\cO_\infty^*$ into 
a bicomplex. Introducing the projector $\vr$ (\ref{r12}) and
the variational operator
$\dl$, one obtains the
variational bicomplex (\ref{7}) of $\cO_\infty^*$. Its
$d_H$- and
$\dl$-cohomology (Theorem
\ref{g90}) has been obtained in several steps
\cite{and,ander,jmp,olv,ijmms,tak2,tul}.

In order to define the variational bicomplex on graded
manifolds (Section 3), let us recall that, by virtue of
Batchelor's theorem \cite{bart}, any graded manifold
$(\gA,X)$ with a body $X$ is isomorphic to the one whose
structure sheaf 
$\gA_Q$ is formed by germs of sections of the exterior product 
\mar{g80}\beq
\w Q^*=\Bbb R\op\oplus_X
Q^*\op\oplus_X\op\w^2 Q^*\op\oplus_X\cdots,
\label{g80}
\eeq 
where $Q^*$ is the dual of some real vector
bundle $Q\to X$. In field models, a vector bundle $Q$ is
usually given from the beginning. Therefore, we consider graded
manifolds
$(X,\gA_Q)$ where 
Batchelor's isomorphism holds. We agree to call 
$(X,\gA_Q)$ the simple graded manifold constructed from $Q$.
Accordingly, $r$-order jets of odd fields are defined as
generating elements of the structure ring of the simple graded
manifold
$(X,\gA_{J^rQ})$ constructed from the jet bundle $J^rQ\to X$ of
$Q$ which is also a vector bundle \cite{book00,mpla}. 
Let $\cC^*_{J^rQ}$ be the bigraded differential algebra
(henceforth BGDA) of $\Bbb Z_2$-graded (or, simply,
graded) exterior forms on the graded manifold
$(X,\gA_{J^rQ})$.  A linear bundle morphism $\pi^r_{r-1}:J^rQ
\to J^{r-1}Q$ yields the corresponding
monomorphism of BGDAs
$\cC^*_{J^{r-1}Q}\to \cC^*_{J^rQ}$ \cite{bart,book00}. Hence,
there is the direct system of BGDAs 
\mar{g205}\beq
\cC^*_Q\ar^{\pi^{1*}_0} \cC^*_{J^1Q}\cdots 
\ar^{\pi^r_{r-1}{}^*}\cC^*_{J^rQ}\ar\cdots,
\label{g205}
\eeq
whose direct limit $\cC^*_\infty$
consists 
of graded exterior forms on graded manifolds
$(X,\gA_{J^rQ})$,
$r\in\Bbb N$, modulo the pull-back identification. 

This definition of odd jets
enables one to describe odd and
even variables (e.g., fields,
ghosts and antifields in BRST theory)
on the same footing. Namely, let $Y\to X$ be an affine bundle 
and $\cP^*_\infty\subset \cO^*_\infty$ the
$C^\infty(X)$-subalgebra of
exterior forms whose coefficients are polynomial in the fiber
coordinates on jet bundles $J^r Y\to X$. This notion is
intrinsic since any element of $\cO^*_\infty$ is an exterior
form on some finite order jet manifold and all jet bundles 
$J^r Y\to X$ are affine. Let us consider the product
$\cS^*_\infty$ of graded algebras $\cC_\infty^*$ and
$\cP^*_\infty$ over their common subalgebra $\cO^*(X)$ of
exterior forms on $X$. It is a BGDA which is
split into the
$\Bbb Z_2$-graded variational bicomplex, analogous to that 
of $\cO^*_\infty$. 

In Section 4, we obtain cohomology of some subcomplexes 
of the variational bicomplex $\cS^*_\infty$ when $X$ is an
arbitrary manifold (Theorem \ref{g96}). They are
the short variational
complex (\ref{g111}) of horizontal (local in the
terminology of
\cite{barn,bran}) graded exterior forms  and the
complex (\ref{g112}) of one-contact graded forms.
For this purpose, one however must: (i) enlarge the BGDA
$\cS^*_\infty$ to the BGDA $\G(\gS^*_\infty)$ of 
graded exterior forms of locally finite jet order, (ii) compute
the cohomology of the corresponding complexes of
$\G(\gS^*_\infty)$, and (iii) prove that this cohomology of 
$\G(\gS^*_\infty)$ coincides with that of $\cS^*_\infty$.
Following this procedure, we show that
cohomology of the complex (\ref{g111}) equals
the de Rham cohomology of $X$, while the complex
(\ref{g112}) is globally exact. 

Note that the exactness of the  
short variational complex (\ref{g111}) on $X=\Bbb R^n$ has been
repeatedly proved \cite{barn,bran,drag}. One has also
considered its subcomplex 
of graded exterior forms whose coefficients are constant on
$\Bbb R^n$. Its $d_H$-cohomology is not trivial
\cite{barn}.

The
exactness of the complex (\ref{g112}) enables us to generalize
the first variational formula and Lagrangian conservation laws
in the calculus of variations on fiber bundles to graded
Lagrangians and  contact supersymmetries of arbitrary order
(Section 5). 

Cohomology of the short
variational complex (\ref{g111}) and its modification
(\ref{g238}) is the main ingredient in a computation of the
iterated cohomology of nilpotent contact supersymmetries.

By analogy with a contact symmetry (Proposition \ref{g62}), an
infinitesimal contact supertransformation or, simply,
 a contact supersymmetry $\up$
is defined as a graded derivation of the
$\Bbb R$-ring
$\cS^0_\infty$ such that the Lie derivative $\bL_\up$
preserves the contact ideal of the
BGDA
$\cS^*_\infty$. 
The BRST transformation
$\up$ (\ref{g130}) in gauge theory on principal bundles 
exemplifies a first order contact supersymmetry such that the
Lie derivative $\bL_\up$  of horizontal graded exterior forms 
is nilpotent. This fact  motivates us to study nilpotent
contact supersymmetries in a general setting. 

The key point is that the Lie derivative
$\bL_\up$ along a contact supersymmetry and the total
differential
$d_H$ mutually commute. When $\bL_\up$ is nilpotent (Lemma
\ref{041}),  we suppose that the
$d_H$-complex $\cS^{0,*}_\infty$ of horizontal graded
exterior forms  is split into the bicomplex $\{S^{k,m}\}$ with
respect to  the 
nilpotent operator
\mar{g240}\beq
\bs_\up\f=(-1)^{|\f|}\bL_\up\f,
\qquad \f\in S^{0,*}_\infty, \label{g240}
\eeq
and the total differential $d_H$.
In the case
of the above mentioned BRST transformation
$\up$ (\ref{g130}), $\bs_\up$ (\ref{g240}) is the BRST operator.
One usually studies the relative cohomology 
$H^{*,*}(\bs_\up/d_H)$ of 
$\bs_\up$ with respect to the 
total differential $d_H$ (see \cite{dub} for the BRST
cohomology modulo the exterior differential $d$).
This cohomology  is not
trivial even when $X=\Bbb R^n$, but it can be related to 
the total
$(\bs_\up+d_H)$-cohomology only in the form degree $n=\di X$.
We consider  the iterated cohomology $H^{*,*}(\bs_\up|d_H)$ of
the bicomplex
$\{S^{k,m}\}$ (Section 6). In the most interesting case of 
form degree $n=\di X$, relative
and iterated cohomology groups 
coincide. They naturally characterize graded Lagrangians
$L\in S^{*,n}$, for which $\up$ is a variational symmetry,
modulo Lie derivatives $\bL_\up\xi$, $\xi\in
S^{0,*}_\infty$, and $d_H$-exact graded exterior forms.

Using the fact that $d_H$-cocycles are represented by exterior
forms on $X$ and that any exterior form on $X$ is
$\bs_\up$-closed, we obtain the iterated cohomology
$H^{*,m<n}(\bs_\up|d_H)$ (Theorem \ref{g250})
and state the relation between the iterated cohomology
$H^{*,n}(\bs_\up|d_H)$ and the total
$(\bs_\up+d_H)$-cohomology of the bicomplex
$S^{*,*}$ (Theorems \ref{04100} and \ref{g251}). Note
that the relative cohomology has also been studied when the
$d_H$-cohomology need not be trivial (\cite{barn}, Section 9.6).
Theorem
\ref{g251}  generalizes this and our result in
\cite{lmp} to an arbitrary nilpotent contact supersymmetry,
when  it may happen that an exterior form on
$X$ is $\bs_\up$-exact.

\section{Lagrangian contact symmetries. Global analysis}

Smooth manifolds throughout are real, finite-dimensional,
Hausdorff, second-countable (hence, paracompact) and connected. 

The projective limit $(J^\infty Y, \pi^\infty_r:J^\infty Y\to
J^rY)$ of the inverse system
(\ref{5.10}) 
is a paracompact Fr\'echet
manifold \cite{tak2}, called the infinite order jet manifold.
 A bundle coordinate atlas $\{(U_Y;x^\la,y^i)\}$ of
$\pi:Y\to X$ yields the coordinate atlas 
\mar{jet1}\beq
\{((\pi^\infty_0)^{-1}(U_Y); x^\la, y^i_\La)\}, \qquad
{y'}^i_{\la+\La}=\frac{\dr x^\m}{\dr x'^\la}d_\m y'^i_\La,
\qquad
0\leq|\La|,
\label{jet1}
\eeq
of $J^\infty Y$, where
$\La=(\la_k...\la_1)$ is a symmetric multi-index,
$\la+\La=(\la\la_k...\la_1)$,  and  
\mar{5.177}\beq
d_\la = \dr_\la + \op\sum_{|\La|\geq 0}
y^i_{\la+\La}\dr_i^\La, \qquad
d_\La=d_{\la_r}\circ\cdots\circ d_{\la_1}, \quad
\La=(\la_r...\la_1), \label{5.177}
\eeq
are the total derivatives. Hereafter, we 
fix an atlas of $Y$ and, consequently, that of $J^\infty Y$
containing a finite number of charts \cite{greub}. 
Restricted to the chart (\ref{jet1}), the GDA
$\cO^*_\infty$ can be written in a
coordinate form; horizontal forms 
$\{dx^\la\}$ and contact one-forms
$\{\th^i_\La=dy^i_\La -y^i_{\la+\La}dx^\la\}$ make up a local
basis for the $\cO^0_\infty$-algebra
$\cO^*_\infty$.

There is the canonical decomposition 
$\cO^*_\infty =\oplus\cO^{k,m}_\infty$
of $\cO^*_\infty$ into $\cO^0_\infty$-modules $\cO^{k,m}_\infty$
of $k$-contact and $m$-horizontal forms
together with the corresponding
projections $h_k:\cO^*_\infty\to \cO^{k,*}_\infty$ and
$h^m:\cO^*_\infty\to \cO^{*,m}_\infty$.
Accordingly, the
exterior differential on $\cO_\infty^*$ is split
into the sum $d=d_H+d_V$ of the total and vertical
differentials 
\be
&& d_H\circ h_k=h_k\circ d\circ h_k, \qquad d_H\circ
h_0=h_0\circ d, \qquad d_H(\f)= dx^\la\w d_\la(\f), \\ 
&& d_V \circ h^m=h^m\circ d\circ h^m, \qquad
d_V(\f)=\th^i_\La \w \dr^\La_i\f, \qquad \f\in\cO^*_\infty.
\ee
One also introduces the $\Bbb R$-module
projector 
\mar{r12}\beq
\vr=\op\sum_{k>0} \frac1k\ol\vr\circ h_k\circ h^n,
\qquad \ol\vr(\f)= \op\sum_{|\La|\geq 0}
(-1)^{\nm\La}\th^i\w [d_\La(\dr^\La_i\rfloor\f)], 
\qquad \f\in \cO^{>0,n}_\infty, \label{r12}
\eeq
of $\cO^*_\infty$ such that
$\vr\circ d_H=0$ and the nilpotent variational operator
$\dl=\vr\circ d_V$ on $\cO^{*,n}_\infty$. Put
$E_k=\vr(\cO^{k,n}_\infty)$. Then the GDA 
$\cO^*_\infty$ is split into the variational
bicomplex
\mar{7}\beq
\begin{array}{ccccrlcrlccrlccrlcrl}
 & &  &  & & \vdots & & & \vdots  & & & 
&\vdots  & & & &
\vdots & &   & \vdots \\ 
& & & & _{d_V} & \put(0,-7){\vector(0,1){14}} & & _{d_V} &
\put(0,-7){\vector(0,1){14}} & &  & _{d_V} &
\put(0,-7){\vector(0,1){14}} & & &  _{d_V} &
\put(0,-7){\vector(0,1){14}}& & _{-\dl} & \put(0,-7){\vector(0,1){14}} \\ 
 &  & 0 & \to & &\cO^{1,0}_\infty &\ar^{d_H} & &
\cO^{1,1}_\infty &
\ar^{d_H} &\cdots  & & \cO^{1,m}_\infty &\ar^{d_H} &\cdots & &
\cO^{1,n}_\infty &\ar^\vr &  & E_1\to  0\\  
& & & & _{d_V} &\put(0,-7){\vector(0,1){14}} & & _{d_V} &
\put(0,-7){\vector(0,1){14}} & & &  _{d_V}
 & \put(0,-7){\vector(0,1){14}} & &  & _{d_V} & \put(0,-7){\vector(0,1){14}}
 & & _{-\dl} & \put(0,-7){\vector(0,1){14}} \\
0 & \to & \Bbb R & \to & & \cO^0_\infty &\ar^{d_H} & &
\cO^{0,1}_\infty &
\ar^{d_H} &\cdots  & &
\cO^{0,m}_\infty & \ar^{d_H} & \cdots & &
\cO^{0,n}_\infty & \equiv &  & \cO^{0,n}_\infty \\
& & & & _{\pi^{\infty*}}& \put(0,-7){\vector(0,1){14}} & & _{\pi^{\infty*}} &
\put(0,-7){\vector(0,1){14}} & & &  _{\pi^{\infty*}}
 & \put(0,-7){\vector(0,1){14}} & &  & _{\pi^{\infty*}} &
\put(0,-7){\vector(0,1){14}} & &  & \\
0 & \to & \Bbb R & \to & & \cO^0(X) &\ar^d & & \cO^1(X) &
\ar^d &\cdots  & &
\cO^m(X) & \ar^d & \cdots & &
\cO^n(X) & \ar^d & 0 &  \\
& & & & &\put(0,-5){\vector(0,1){10}} & & &
\put(0,-5){\vector(0,1){10}} & & & 
 & \put(0,-5){\vector(0,1){10}} & & &   &
\put(0,-5){\vector(0,1){10}} & &  & \\
& & & & &0 & &  & 0 & & & & 0 & & & & 0 & &  & 
\end{array}
\label{7}
\eeq
Its
cohomology has been obtained in several  steps (see the outline
of proof of Theorem
\ref{g90} in Appendix A).  

\begin{theo} \label{g90} \mar{g90} 
(i) The second row from the bottom and the last column of this
bicomplex make up the variational complex
\mar{b317}\beq
0\to\Bbb R\to \cO^0_\infty
\ar^{d_H}\cO^{0,1}_\infty\cdots  
\op\longrightarrow^{d_H} 
\cO^{0,n}_\infty  \op\longrightarrow^\dl E_1 
\op\longrightarrow^\dl 
E_2 \ar \cdots\,. \label{b317} 
\eeq
Its cohomology is isomorphic to the de Rham cohomology of the
fiber bundle
$Y$. (ii) The rows of contact forms of the bicomplex (\ref{7})
are exact sequences.
\end{theo}

One can think of the elements
\be
L=\cL\om\in \cO^{0,n}_\infty, 
\qquad \dl L=\op\sum_{|\La|\geq
0}(-1)^{|\La|}d_\La(\dr^\La_i \cL)\th^i\w\om\in E_1, \qquad
 \om=dx^1\w\cdots\w dx^n,
\ee
of the variational complex (\ref{b317}) as being
a finite order Lagrangian and its
Euler--Lagrange operator, respectively.

\begin{cor} \label{g93} \mar{g93}
The exactness of the row of one-contact forms of
the variational bicomplex (\ref{7}) at the term
$\cO^{1,n}_\infty$ relative to the 
projector $\vr$ provides 
the $\Bbb R$-module decomposition
\be
\cO^{1,n}_\infty=E_1\oplus d_H(\cO^{1,n-1}_\infty)
\ee
and, given a Lagrangian $L\in \cO^{0,n}_\infty$, the
corresponding decomposition
\mar{+421}\beq
dL=\dl L-d_H\Xi. \label{+421}
\eeq
\end{cor}

The form $\Xi$ in the decomposition (\ref{+421}) is not
uniquely defined. It reads
\be
\Xi=\op\sum_{s=0}F^{\la\nu_s\ldots\nu_1}_i
\th^i_{\nu_s\ldots\nu_1}\w\om_\la, \quad 
F_i^{\nu_k\ldots\nu_1}=
\dr_i^{\nu_k\ldots\nu_1}\cL-d_\la F_i^{\la\nu_k\ldots\nu_1}
+h_i^{\nu_k\ldots\nu_1}, \quad 
\om_\la=\dr_\la\rfloor\om,
\ee
where local functions $h\in\cO^0_\infty$ obey the relations
$h^\nu_i=0$,
$h_i^{(\nu_k\nu_{k-1})\ldots\nu_1}=0$. It follows that
$\Xi_L=\Xi +L$ is a Lepagean equivalent of a finite order
Lagrangian $L$ \cite{got}.

The decomposition (\ref{+421})
leads to the first variational formula (\ref{g8}) and
the Lagrangian conservation law (\ref{g32}) as follows.

A derivation $\up\in\gd\cO^0_\infty$ of the
$\Bbb R$-ring $\cO^0_\infty$ such that the Lie derivative
$\bL_\up$ preserves the contact ideal of the GDA
$\cO^*_\infty$ (i.e., the Lie derivative $\bL_\up$ of a
contact form
 is a contact form) is called an infinitesimal contact
transformation or, simply, a contact symmetry (by  analogy with
$\cC$-transformations in
\cite{kras} though $\up$ need not come from a morphism of
$J^\infty Y$). Proposition \ref{g62} below shows that,
restricted to a coordinate chart (\ref{jet1}) and a GDA
$\cO^*_r$ of finite jet order, any contact symmetry $\up$ is the
jet prolongation of a generalized vector field in \cite{olv}. 

\begin{prop} \label{g62} \mar{g62}
(i) The derivation module $\gd\cO^0_\infty$ is isomorphic to
the 
$\cO^0_\infty$-dual $(\cO^1_\infty)^*$ of the module of
one-forms $\cO^1_\infty$.
(ii) Relative to an atlas
(\ref{jet1}), a derivation $\up\in\gd\cO^0_\infty$ is
given by the expression
\mar{g3}\beq
\up=\up^\la \dr_\la + \up^i\dr_i +
\op\sum_{|\La|>0}\up^i_\La
\dr^\La_i, \label{g3}  
\eeq
where $\up^\la$, $\up^i$, $\up^i_\La$ are local
smooth functions of finite jet order obeying the transformation
law
\mar{g71}\beq
\up'^\la=\frac{\dr x'^\la}{\dr x^\m}\up^\m, \qquad
\up'^i=\frac{\dr y'^i}{\dr y^j}\up^j + \frac{\dr y'^i}{\dr
x^\m}\up^\m, \qquad 
\up'^i_\La=\op\sum_{|\Si|\leq|\La|}\frac{\dr y'^i_\La}{\dr
y^j_\Si}\up^j_\Si +
\frac{\dr y'^i_\La}{\dr x^\m}\up^\m. \label{g71} 
\eeq
(iii) A derivation $\up$ (\ref{g3}) is a contact
symmetry iff 
\mar{g4}\beq
\up^i_\La=d_\La(\up^i-y^i_\m\up^\m)+y^i_{\m+\La}\up^\m, \qquad
0<|\La|.
\label{g4}
\eeq
\end{prop}

\begin{proof} (i)
At first, let us show that $\cO^*_\infty$ is generated by
elements $df$,
$f\in \cO^0_\infty$. It suffices to justify that any
element of $\cO^1_\infty$ is a finite $\cO^0_\infty$-linear
combination of elements $df$,
$f\in \cO^0_\infty$. Indeed, every
$\f\in\cO^1_\infty$ is an exterior form on some finite order
jet manifold $J^rY$.  By
virtue of the Serre--Swan theorem extended to non-compact manifolds \cite{ren,ss}, the
$C^\infty(J^rY)$-module $\cO^*_r$ of one-forms on $J^rY$ is
a projective module of finite rank, i.e., $\f$ is
represented by a finite $C^\infty(J^rY)$-linear combination of
elements $df$,
$f\in C^\infty(J^rY)\subset
\cO^0_\infty$.  Any element $\Phi\in (\cO^1_\infty)^*$ yields a
derivation
$\up_\Phi(f)=\Phi(df)$ of the $\Bbb R$-ring $\cO^0_\infty$.
Since the module
$\cO^1_\infty$ is generated by elements $df$, $f\in
\cO^0_\infty$, different elements of $(\cO^1_\infty)^*$
provide different derivations of $\cO^0_\infty$, i.e., there
is a monomorphism  $(\cO^1_\infty)^*\to \gd\cO^0_\infty$. 
By
the same formula, any derivation $\up\in \gd\cO^0_\infty$
sends $df\mapsto \up(f)$ and, since $\cO^0_\infty$ is generated
by elements $df$, it defines a morphism 
$\Phi_\up:\cO^1_\infty\to
\cO^0_\infty$. Moreover, different derivations $\up$ provide
different morphisms $\Phi_\up$. Thus, we have a monomorphism
and, consequently, an isomorphism $\gd\cO^0_\infty\to 
(\cO^1_\infty)^*$. 

(ii) Restricted to a coordinate chart (\ref{jet1}), 
$\cO^1_\infty$ is a free $\cO^0_\infty$-module
generated by the
exterior forms $dx^\la$, $\th^i_\La$. Then
$\gd\cO^0_\infty=(\cO^1_\infty)^*$ restricted to this chart 
consists of elements (\ref{g3}), where $\dr_\la$, $\dr^\La_i$
are the duals of $dx^\la$, $\th^i_\La$. The
transformation rule (\ref{g71}) results from the transition
functions (\ref{jet1}). 
The interior product $\up\rfloor\f$ 
and the Lie derivative $\bL_\up\f$,
$\f\in\cO^*_\infty$, obey  
the standard formulae. 
Restricted to a coordinate chart, the Lie derivative $\bL_\up$
sends each finite jet order GDA $\cO^*_r$ to another finite jet
order GDA $\cO^*_s$. Since the atlas (\ref{jet1}) is
finite, $\bL_\up\f$ preserves $\cO^*_\infty$.

(iii) 
The expression (\ref{g4}) results from a direct
computation  similar to that of the first part
of B\"acklund's theorem \cite{ibr}. One can then 
justify that local functions (\ref{g4}) satisfy the 
transformation law (\ref{g71}).
\end{proof}

Any contact symmetry admits the horizontal
splitting
\mar{g5}\beq
\up=\up_H +\up_V=\up^\la d_\la + (\vt^i\dr_i +
\op\sum_{|\La|>0} d_\La \vt^i\dr_i^\La), \qquad \vt^i=
\up^i-y^i_\m\up^\m,
\label{g5}
\eeq
 relative to the canonical connection
$\nabla=dx^\la\ot d_\la$ on the $C^\infty(X)$-ring
$\cO^0_\infty$ \cite{book00}.

\begin{lem} \label{0410} \mar{0410}
Any
vertical contact symmetry
$\up=\up_V$ obeys the relations
\mar{g6,'}\ben
&& \up\rfloor d_H\f=-d_H(\up\rfloor\f), \label{g6}\\
&& \bL_\up(d_H\f)=d_H(\bL_\up\f), \qquad \f\in\cO^*_\infty.
\label{g6'}
\een 
\end{lem}

\begin{proof} It is easily justified that, if $\f$ and
$\f'$ satisfy the relation (\ref{g6}), then $\f\w\f'$
does well. Then it suffices to prove the relation (\ref{g6})
when $\f$ is a function and $\f=\th^i_\La$. The
result follows from the equalities
\mar{0480,'}\ben
&& \up\rfloor \th^i_\La=\up^i_\La, \qquad 
d_H(\up^i_\La)=\up^i_{\la+\La}dx^\la, \qquad
d_H\th^i_\la=dx^\la\w\th^i_{\la+\La}, \label{0480}\\
&& d_\la\circ v^i_\La\dr_i^\La= v^i_\La\dr_i^\La \circ d_\la.
\label{0480'}
\een
The relation (\ref{g6'}) is a corollary of the equality
(\ref{g6}).
\end{proof}

\begin{prop}  \label{g75} \mar{g75}
Given a  Lagrangian $L=\cL\om\in\cO^{0,n}_\infty$, its 
Lie derivative $\bL_\up L$
along a contact
symmetry
$\up$ (\ref{g5}) fulfils the first variational formula
\mar{g8}\beq
\bL_\up L= \up_V\rfloor\dl L +d_H(h_0(\up\rfloor\Xi_L)) 
+\cL d_V (\up_H\rfloor\om), \label{g8}
\eeq
where $\Xi_L$ is a Lepagean equivalent, e.g.,
a Poincar\'e--Cartan form of $L$.
\end{prop}

\begin{proof}
The formula (\ref{g8}) comes from the splitting
(\ref{+421}) and the relation (\ref{g6}) as follows:  
\be
&& \bL_\up L=\up\rfloor dL + d(\up\rfloor L)
=[\up_V\rfloor dL -d_V\cL\w \up_H\rfloor\om]
+[d_H(\up_H\rfloor L) + d_V(\cL
\up_H\rfloor\om)]= \\
&& \quad \up_V\rfloor dL + d_H(\up_H\rfloor L) +\cL d_V
(\up_H\rfloor\om)=   \up_V\rfloor\dl L -\up_V\rfloor
d_H\Xi + d_H(\up_H\rfloor L) +\cL d_V (\up_H\rfloor\om)
=  \\
&& \quad \up_V\rfloor\dl L +d_H(\up_V\rfloor\Xi +
\up_H\rfloor L) +\cL d_V (\up_H\rfloor\om), 
\ee
where $\up_V\rfloor\Xi=h_0(\up\rfloor\Xi)$ since $\Xi$ is a
one-contact form,
$\up_H\rfloor L=h_0(\up\rfloor L)$, and
$\Xi_L=\Xi+L$.
\end{proof}

Let $\up$ be a variational symmetry of $L$ (in the
terminology of \cite{olv}), i.e.,
$\bL_\up L=d_H\si$, $\si\in
\cO^{0,n-1}_\infty$.
By virtue of the expression (\ref{g8}), this condition implies
that $\up$ is projected onto $X$.
Then the first variational formula (\ref{g8}) restricted to
Ker$\,\dl L$ leads to the weak conservation law
\mar{g32}\beq
0\ap d_H(h_0(\up\rfloor\Xi_L)-\si). \label{g32}
\eeq

\section{$\Bbb Z_2$-Graded variational bicomplex}

Let $(X,\gA_Q)$ be the simple graded manifold constructed from
a vector bundle $Q\to X$ of fiber dimension $m$.
Its structure ring
$\cA_Q$ of sections of $\gA_Q$ consists of sections of the
exterior bundle (\ref{g80}) called graded functions. Given
bundle coordinates
$(x^\la,q^a)$ on
$Q$ with transition functions
$q'^a=\rho^a_b q^b$, let
$\{c^a\}$ be the corresponding fiber bases for
$Q^*\to X$, together with transition functions
$c'^a=\rho^a_bc^b$. Then $(x^\la, c^a)$ is called the local
basis for the graded manifold $(X,\gA_Q)$ \cite{bart,book00}.
With respect to this basis, graded functions read 
\be
f=\op\sum_{k=0}^m \frac1{k!}f_{a_1\ldots
a_k}c^{a_1}\cdots c^{a_k}, 
\ee
where $f_{a_1\cdots
a_k}$ are local smooth real functions on $X$.

Given a graded manifold $(X,\gA_Q)$, by the sheaf $\gd\gA_Q$ of
graded derivations of $\gA_Q$ is meant a subsheaf of
endomorphisms of the structure sheaf $\gA_Q$ such that any
section $u$ of $\gd\gA_Q$ over an open subset $U\subset X$ is a
$\Bbb Z_2$-graded derivation of the $\Bbb Z_2$-graded ring
$\cA_Q(U)$ of graded functions on $U$, i.e.,
\be
u(ff')=u(f)f'+(-1)^{[u][f]}fu (f'), 
\qquad f,f'\in
\cA_Q(U),                                                                                           
\ee
where $[.]$
denotes the Grassmann parity.
One can show that sections of $\gd\gA_Q$ over $U$ exhaust all
$\Bbb Z_2$-graded derivations of the ring $\cA_Q(U)$
\cite{bart}. Let $\gd\cA_Q$ be the Lie
superalgebra of $\Bbb Z_2$-graded derivations
of the $\Bbb R$-ring $\cA_Q$. 
Its elements are
called $\Bbb Z_2$-graded (or, simply, graded) vector fields on
$(X,\gA_Q)$. Due to the canonical splitting
$VQ= Q\times Q$, the vertical tangent bundle 
$VQ\to Q$ of $Q\to X$ can be provided with the fiber bases 
$\{\dr_a\}$ which is the dual of 
$\{c^a\}$. Then
a graded vector field takes the local form
$u= u^\la\dr_\la + u^a\dr_a$,
where $u^\la, u^a$ are local graded functions. It
acts on
$\cA_Q$ by the rule
\mar{cmp50'}\beq
u(f_{a\ldots b}c^a\cdots c^b)=u^\la\dr_\la(f_{a\ldots b})c^a\cdots c^b +u^d
f_{a\ldots b}\dr_d\rfloor (c^a\cdots c^b). \label{cmp50'}
\eeq
This rule implies the corresponding transformation law 
\be
u'^\la =u^\la, \qquad u'^a=\rho^a_ju^j +
u^\la\dr_\la(\rho^a_j)c^j. 
\ee
Then one can show \cite{book00,ijmp} that graded vector fields 
on a simple graded manifold can be represented
by sections of the vector bundle
$\cV_Q\to X$ which is locally isomorphic to the vector bundle
$\w Q^*\ot_X(Q\oplus_X TX)$,
and is equipped with the bundle coordinates
$(\dot x^\la_{a_1\ldots a_k},v^i_{b_1\ldots b_k})$,
$k=0,\ldots,m$, together with the transition functions
\be
&& \dot x'^\la_{i_1\ldots i_k}=\rho^{-1}{}_{i_1}^{a_1}\cdots
\rho^{-1}{}_{i_k}^{a_k}\dot x^\la_{a_1\ldots a_k},\\
&& v'^i_{j_1\ldots j_k}=\rho^{-1}{}_{j_1}^{b_1}\cdots
\rho^{-1}{}_{j_k}^{b_k}\left[\rho^i_jv^j_{b_1\ldots b_k}+ \frac{k!}{(k-1)!} 
\dot x^\la_{b_1\ldots b_{k-1}}\dr_\la\rho^i_{b_k}\right].
\ee

Using this fact, we can introduce graded exterior forms on the
simple graded manifold $(X,\gA_Q)$ as sections of the 
exterior bundle $\op\w\cV^*_Q$, where 
$\cV^*_Q\to  X$ is the pointwise $\w Q^*$-dual of $\cV_Q$.
Relative to the dual bases $\{dx^\la\}$ for $T^*X$ and
$\{dc^b\}$ for $Q^*$, graded
one-forms read 
\be
\f=\f_\la dx^\la + \f_adc^a,\qquad \f'_a=\rho^{-1}{}_a^b\f_b,
\qquad
\f'_\la=\f_\la +\rho^{-1}{}_a^b\dr_\la(\rho^a_j)\f_bc^j.
\ee
The duality morphism is given by the interior product 
\be
u\rfloor \f=u^\la\f_\la + (-1)^{[\f_a]}u^a\f_a. 
\ee
Graded exterior forms constitute the BGDA $\cC^*_Q$ with
respect to the bigraded exterior product $\w$ and the 
exterior differential $d$. The standard formulae of a BGDA hold.

Since the jet bundle $J^rQ\to X$ of a vector bundle $Q\to X$
is a vector bundle, let us consider the simple graded manifold
$(X,\gA_{J^rQ})$ constructed from 
$J^rQ\to X$. Its local basis is $\{x^\la,c^a_\La\}$, 
$0\leq |\La|\leq r$,
together with the transition functions
\mar{+471}\beq
c'^a_{\la +\La}=d_\la(\rho^a_j c^j_\La),
\qquad 
d_\la=\dr_\la + \op\sum_{|\La|<r}c^a_{\la+\La}
\dr_a^\La, \label{+471}
\eeq 
where $\dr_a^\La$ are the duals of $c^a_\La$.  
Let $\cC^*_{J^rQ}$ be the BGDA of graded
exterior forms on the graded manifold $(X,\gA_{J^rQ})$.
 The direct limit
$\cC^*_\infty$ of the direct system (\ref{g205}) inherits
the BGDA operations intertwined by the
monomorphisms
$\pi^r_{r-1}{}^*$. It is locally a free
$C^\infty(X)$-algebra countably generated by the elements 
$(1, c^a_\La, dx^\la,\th^a_\La=dc^a_\La -c^a_{\la
+\La}dx^\la)$, $0\leq|\La|$.

It should be emphasized that, in contrast with the GDA
$\cO^*_\infty$, the BGDA $\cC^*_\infty$ consists
of sections of sheaves over $X$. In order to regard these
algebras  on the same footing,  let $Y\to X$ hereafter be an
affine bundle. Then one can show that the GDA 
$\cO^*_\infty$ is an algebra  of
sections of some sheaf over $X$ (see Appendix B). Let us 
consider the above mentioned polynomial subalgebra
$\cP^*_\infty$ of
$\cO^*_\infty$ and the product 
$\cC_\infty^*\w\cP^*_\infty$ of
graded algebras $\cC_\infty^*$ and $\cP^*_\infty$ over their
common graded subalgebra $\cO^*(X)$ of exterior forms on $X$. It
consists of the elements 
\be
\op\sum_i \psi_i\ot\f_i, \qquad \op\sum_i \f_i\ot\psi_i, \qquad
\psi\in \cC^*_\infty, \qquad \f\in \cP^*_\infty,
\ee
of the tensor products $\cC_\infty^*\ot
\cP^*_\infty$ and
$\cP_\infty^*\ot \cC^*_\infty$ of the
$C^\infty(X)$-modules 
$\cC_\infty^*$ and $\cP^*_\infty$ which 
are subject to the
commutation relations
\mar{0442}\ben
&&\psi\ot\f=(-1)^{|\psi||\f|}\f\ot\psi, \qquad
\psi\in \cC^*_\infty, \qquad \f\in \cP^*_\infty, \label{0442}\\
&& (\psi\w\si)\ot\f=\psi\ot(\si\w\f),
\qquad \si\in
\cO^*(X), \nonumber
\een
and the multiplication
\mar{0440}\beq
(\psi\ot\f)\w(\psi'\ot\f'):=(-1)^{|\psi'||\f|}(\psi\w\psi')\ot
(\f\w\f'). \label{0440}
\eeq
Elements $\psi\ot\f$ are endowed
with the total form degree
$|\psi|+|\f|$ and the total Grassmann
parity $[\psi]$. Then the multiplication (\ref{0440}) obeys
the relation  
\be
\vf\w\vf' =(-1)^{|\vf||\vf'| +[\vf][\vf']}\vf'\w \vf, \qquad
\vf,\vf'\in \cS^*_\infty,  
\ee
and makes $\cC_\infty^*\w\cP^*_\infty$ into a
bigraded $C^\infty(X)$-algebra $\cS^*_\infty$, where the
asterisk means the total form degree. Due to the algebra
monomorphisms
\be
\cC^*_\infty\ni\psi \to \psi\ot 1=1\ot\psi\in \cS^*_\infty,
\qquad 
\cP^*_\infty\ni\f \to \f\ot 1=1\ot\f\in \cS^*_\infty,
\ee
one can think of $\cS^*_\infty$ as being an algebra
generated by elements of $\cC_\infty^*$ and $\cP^*_\infty$. 
For instance, elements
of the ring
$S^0_\infty$ are polynomials of
$c^a_\La$ and $y^i_\La$ with coefficients in $C^\infty(X)$. 

Let us provide $\cS^*_\infty$ with the exterior differential
\mar{0441}\beq
d(\psi\ot\f):=(d_\cC\psi)\ot\f +(-1)^{|\psi|}\psi\ot(d_\cP\f),
\qquad 
\psi\in \cC^*_\infty, \qquad \f\in \cP^*_\infty, \label{0441}
\eeq
where $d_\cC$ and $d_\cP$ are exterior differentials on the
differential algebras $\cC^*_\infty$ and $\cP^*_\infty$,
respectively. We obtain at once from the relation (\ref{0442})
that
\be
d(\f\ot\psi)=(d_\cP\f)\ot\psi
+(-1)^{|\f|}\f\ot(d_\cC\psi),
\qquad 
\psi\in \cC^*_\infty, \qquad \f\in \cP^*_\infty.
\ee
The exterior
differential $d$ (\ref{0441}) is nilpotent. It obeys the
equalities
\be
 d(\vf\w\vf')= d\vf\w\vf' +(-1)^{|\vf|}\vf\w d\vf', \qquad
\vf,\vf'\in \cS^*_\infty,
\ee
and makes $\cS^*_\infty$ into a BGDA, which
is locally generated by the elements 
\be
(1, c^a_\La, y^i_\La,
dx^\la,\th^a_\La=dc^a_\La-c^a_{\la+\La}dx^\la,\th^i_\La=
dy^i_\La-y^i_{\la+\La}dx^\la), \qquad |\La|\geq 0.
\ee

Hereafter, let the
collective symbols
$s^A_\La$ and $\th^A_\La$ stand both for even and odd
generating elements
$c^a_\La$, $y^i_\La$, $\th^a_\La$, $\th^i_\La$ of the
$C^\infty(X)$-algebra
$\cS^*_\infty$ which, thus, is locally generated by $(1,s^A_\La,
dx^\la, \th^A_\La)$, $|\La|\geq 0$. We agree to call elements
of $\cS^*_\infty$ the graded exterior forms on $X$.

Similarly to
$\cO^*_\infty$, the BGDA
$\cS^*_\infty$ is decomposed  into
$\cS^0_\infty$-modules $\cS^{k,r}_\infty$ of
$k$-contact and
$r$-horizontal graded forms together with the corresponding
projections $h_k$ and $h^r$. Accordingly, the exterior 
differential
$d$ (\ref{0441}) on 
$\cS^*_\infty$ is split into the sum $d=d_H+d_V$ of 
the total and vertical differentials 
\be
d_H(\f)=dx^\la\w d_\la(\f), \qquad d_V(\f)=\th^A_\La\w\dr^\La_A
\f, \qquad \f\in \cS^*_\infty.
\ee
The projection
endomorphism $\vr$ of $\cS^*_\infty$ is given by the expression
\be
\vr=\op\sum_{k>0} \frac1k\ol\vr\circ h_k\circ h^n,
\qquad \ol\vr(\f)= \op\sum_{|\La|\geq 0}
(-1)^{\nm\La}\th^A\w [d_\La(\dr^\La_A\rfloor\f)], 
\qquad \f\in \cS^{>0,n}_\infty,
\ee
similar to (\ref{r12}). The graded variational operator
$\dl=\vr\circ d$ is introduced. Then the BGDA
$\cS^*_\infty$ is split into the $\Bbb Z_2$-graded variational
bicomplex
\mar{0453}\beq
(\cO^*(X), \cS^{*,*}_\infty,
E_k=\vr(\cS^{k,n}_\infty);d,d_H,d_V,\vr,\dl),
\label{0453}
\eeq
analogous to the variational bicomplex (\ref{7}). 

\section{Cohomology of $\Bbb Z_2$-graded complexes}

We aim to study the cohomology of the 
short variational complex
\mar{g111}\beq
0\ar \Bbb R\ar
\cS^0_\infty\ar^{d_H}\cS^{0,1}_\infty \cdots
\ar^{d_H} \cS^{0,n}_\infty\ar^\dl E_1 
\label{g111}  
\eeq
and the complex of one-contact graded forms
\mar{g112}\beq
 0\to \cS^{1,0}_\infty\ar^{d_H} \cS^{1,1}_\infty\cdots
\ar^{d_H}\cS^{1,n}_\infty\ar^\vr E_1\to 0
\label{g112}
\eeq
of the BGDA $\cS^*_\infty$.
One can think of the elements 
\be
L=\cL\om\in \cS^{0,n}_\infty, \qquad 
\dl (L)= \op\sum_{|\La|\geq 0}
 (-1)^{|\La|}\th^A\w d_\La (\dr^\La_A L)\in E_1
\ee
of the complexes (\ref{g111}) -- (\ref{g112}) as being a graded
Lagrangian and its Euler--Lagrange operator, respectively.

\begin{theo} \label{g96} \mar{g96} The cohomology of
the complex (\ref{g111}) equals the de
Rham cohomology $H^*(X)$ of 
$X$. The complex (\ref{g112}) is exact.
\end{theo}

The proof of
Theorem \ref{g96} follows
the scheme of the proof of Theorem \ref{g90}, but all sheaves
are sheaves over $X$. The proof falls into the three steps. 

(i) We start by
showing that the complexes (\ref{g111}) -- (\ref{g112}) are
locally exact.

\begin{lem} \label{0465} \mar{0465}
The complex (\ref{g111}) on
$X=\Bbb R^n$ is exact.
\end{lem}

Referring to  \cite{barn}, Theorems 4.1
-- 4.2, for the proof, we summarize a few formulae 
quoted in the sequel. Any horizontal graded form $\f\in
\cS^{0,*}_\infty$ admits the decomposition 
\mar{0471}\beq
\f=\f_0 + \wt\f, \qquad \wt\f=
\op\int^1_0\frac{d\la}{\la}\op\sum_{|\La|\geq 0}s^A_\La
\dr^\La_A\f,
\label{0471}
\eeq
where $\f_0$ is an exterior form on $\Bbb R^n$.
Let $\f\in \cS^{0,m<n}_\infty$ be
$d_H$-closed.  Then its component $\f_0$ (\ref{0471}) is an
exact exterior form on
$\Bbb R^n$  and $\wt\f=d_H\xi$, where $\xi$
is given by the following expressions. Let us introduce the
operator
\mar{0470}\beq
D^{+\nu}\wt\f=\op\int^1_0\frac{d\la}{\la}\sum_{k\geq 0}
k\dl^\nu_{(\m_1}\dl^{\al_1}_{\m_2}\cdots\dl^{\al_{k-1}}_{\m_k)}
\la s^A_{(\al_1\ldots\al_{k-1})}
\dr_A^{\m_1\ldots\m_k}\wt\f(x^\m,\la
s^A_\La, dx^\m). \label{0470}
\eeq
The relation
$[D^{+\nu},d_\m]\wt\f=\dl^\nu_\m\wt\f$ holds, and leads to the
desired expression
\mar{0473}\beq
\xi=\op\sum_{k=0}\frac{(n-m-1)!}{(n-m+k)!}D^{+\nu} P_k
\dr_\nu\rfloor\wt\f, \qquad P_0=1, \quad
 P_k=d_{\nu_1}\cdots d_{\nu_k}D^{+\nu_1}\cdots
D^{+\nu_k}.
 \label{0473} 
\eeq 
Now let $\f\in \cS^{0,m<n}_\infty$ be a graded density such
that $\dl\f=0$. Then its component $\f_0$
(\ref{0471}) is an exact form on $\Bbb R^n$ and $\wt\f=d_H\xi$,
where
$\xi$ is given by the expression
\mar{0474}\beq
\xi=\op\sum_{|\La|\geq
0}\op\sum_{\Si+\Xi=\La}(-1)^{|\Si|}s^A_\Xi
d_\Si\dr^{\m+\La}_A\wt\f\om_\m. \label{0474}
\eeq

\begin{rem} \label{0476} \mar{0476}
Since elements of $\cS^*_\infty$ are polynomials in $s^A_\La$,
the sum in the expression (\ref{0473}) is finite. However,
the expression (\ref{0473}) contains a $d_H$-exact summand
which prevents its extension to  
$\cO^*_\infty$. 
In this respect, we also quote the
homotopy operator (5.107) in \cite{olv} which leads to
the expression
\mar{0477,'}\ben
&&\xi=\op\int_0^1 I(\f)(x^\m,\la s^A_\La,
dx^\m)\frac{d\la}{\la}, \label{0477}\\
&& I(\f)=\op\sum_{|\La|\geq 0}\op\sum_\m
\frac{\La_\m+1}{n-m+|\La|+1} d_\La[ \op\sum_{|\Xi|\geq 0}
(-1)^\Xi \frac{(\m+\La+\Xi)!}{(\m+\La)!\Xi!}s^A d_\Xi
\dr_A^{\m+\La+\Xi}(\dr_\m\rfloor \f)], \label{0477'}
\een
where $\La!=\La_{\m_1}!\cdots \La_{\m_n}!$ and $\La_\m$
denotes the number of occurrences of the index $\m$ in $\La$.
The graded forms (\ref{0474}) and (\ref{0477}) differ in a
$d_H$-exact graded form.
\end{rem}

\begin{lem} \label{g220} \mar{g220}
The complex (\ref{g112}) on $X=\Bbb R^n$ is exact.
\end{lem}

\begin{proof}
The fact that a $d_H$-closed graded $(1,m)$-form
$\f\in \cS^{1,m<n}_\infty$ is $d_H$-exact is derived from  
Lemma \ref{0465} as follows. We write 
\mar{0445}\beq
\f=\sum\f_A^\La\w \th^A_\La, \label{0445}
\eeq
where $\f_A^\La\in
\cS^{0,m}_\infty$ are horizontal graded $m$-forms. Let
us introduce additional variables $\ol s^A_\La$ of the same 
Grassmann parity as $s^A_\La$. Then one can 
associate to each graded $(1,m)$-form $\f$ (\ref{0445})
a unique horizontal graded
$m$-form 
\mar{0446}\beq
\ol\f=\sum\f_A^\La\ol s^A_\La, \label{0446}
\eeq
whose coefficients are linear in the variables $\ol s^A_\La$,
and {\it vice versa}.  Let us consider the modified total
differential
\be
\ol d_H=d_H + dx^\la\w \op\sum_{|\La|>0}\ol
s^A_{\la+\La}\ol\dr_A^\La, 
\ee
acting on graded forms (\ref{0446}), where $\ol\dr^\La_A$ is
the dual of $d\ol s^A_\La$. Comparing the equality 
$\ol d_H\ol s^A_\La=dx^\la s^A_{\la+\La}$ and the last equality
(\ref{0480}), one can  easily justify that
$\ol{d_H\f}=\ol d_H\ol\f$. Let a 
graded $(1,m)$-form $\f$ (\ref{0445}) be $d_H$-closed. Then 
the associated horizontal graded $m$-form $\ol \f$
(\ref{0446}) is
$\ol d_H$-closed and, by virtue of Lemma \ref{0465}, 
it is $\ol d_H$-exact, i.e., 
$\ol \f= \ol d_H \ol\xi$, where
$\ol\xi$ is a horizontal graded
$(m-1)$-form given by the expression (\ref{0473}) depending  on
additional variables $\ol s^A_\La$. A glance at this
expression shows that, since $\ol\f$ is linear in the variables 
$\ol s^A_\La$, so is $\ol\xi=\sum\xi_A^\La\ol s^A_\La$. It
follows that
$\f=d_H\xi$ where
$\xi=\sum\xi_A^\La\w \th^A_\La$. It remains to prove the
exactness of the complex (\ref{g112}) at the last term $E_1$.
If 
\be
\vr(\si)=\op\sum_{|\La|\geq 0}(-1)^{|\La|}\th^A\w
[d_\La(\dr_A^\La\rfloor\si)]= \op\sum_{|\La|\geq 0}
(-1)^{|\La|}\th^A\w
[d_\La\si_A^\La]\om=0, \qquad  \si\in
\cS^{1,n}_\infty,
\ee
a direct computation gives 
\mar{0449}\beq
\si=d_H\xi,\qquad  \xi=-\op\sum_{|\La|\geq
0}\op\sum_{\Si+\Xi=\La} (-1)^{|\Si|}\th^A_{\Xi}\w
d_\Si\si^{\m+\La}_A \om_\m. \label{0449}
\eeq
\end{proof}

\begin{rem}
The proof of Lemma \ref{g220} fails to be extended to 
complexes of higher contact forms because
the products $\th^A_\La\w\th^B_\Si$ and $s^A_\La s^B_\Si$ obey
different commutation rules.
\end{rem} 

(ii) Let us associate to each open subset $U\subset X$ the BGDA 
 $\cS^*_U$ of elements of the
$C^\infty(X)$-algebra $\cS^*_\infty$ whose coefficients
are restricted to
$U$. These algebras make up a presheaf over $X$. Let
$\gS^*_\infty$ be the sheaf of germs of this
presheaf and
$\G(\gS^*_\infty)$ its structure module of sections. One can
show that $\gS^*_\infty$ inherits the  variational bicomplex
operations, and $\G(\gS^*_\infty)$ does so (see Appendix C). 
For short, we say that $\G(\gS^*_\infty)$ consists of
polynomials in $s^a_\La$, $ds^a_\La$ of
locally bounded jet order $|\La|$. There is the monomorphism
$\cS^*_\infty\to\G(\gS^*_\infty)$. 
Let us consider the
complexes of sheaves  
\mar{g114,5}\ben
&& 0\ar \Bbb R\ar
\gS^0_\infty\ar^{d_H}\gS^{0,1}_\infty \cdots
\ar^{d_H} \gS^{0,n}_\infty\ar^\dl \gE_1, \qquad
\gE_1=\vr(\gS^{1,n}_\infty),
\label{g114}\\
&& 0\to \gS^{1,0}_\infty\ar^{d_H} \gS^{1,1}_\infty\cdots
\ar^{d_H}\gS^{1,n}_\infty\ar^\vr \gE_1\to 0 
 \label{g115}
\een
over $X$ and the complexes of their structure modules
\mar{g117,8}\ben
&& 0\ar \Bbb R\ar
\G(\gS^0_\infty)\ar^{d_H}\G(\gS^{0,1}_\infty) \cdots
\ar^{d_H} \G(\gS^{0,n}_\infty)\ar^\dl  \G(\gE_1), 
\label{g117}\\
&& 0\to \G(\gS^{1,0}_\infty)\ar^{d_H} \G(\gS^{1,1}_\infty)\cdots
\ar^{d_H}\G(\gS^{1,n}_\infty)\ar^\vr \G(\gE_1)\to 0.
\label{g118}
\een
By virtue of Lemmas \ref{0465} -- \ref{g220} and Theorem
\ref{0460}, the complexes of sheaves (\ref{g114}) --
(\ref{g115}) are exact.
The terms $\gS^{*,*}_\infty$ of the complexes (\ref{g114}) --
(\ref{g115}) are sheaves of
$C^\infty(X)$-modules. Therefore, they are fine and,
consequently, acyclic. By virtue of Theorem \ref{+132} (see
Appendix D), the cohomology of the complex (\ref{g117}) equals
the cohomology of $X$ with coefficients in the constant sheaf
$\Bbb R$, i.e., the de Rham cohomology
$H^*(X)$ of $X$, whereas the complex (\ref{g118}) is
globally exact. 

(iii) It remains to prove the following.

\begin{prop} \label{g239} \mar{g239}
Cohomology of the complexes (\ref{g111})
-- (\ref{g112}) equals that of the complexes (\ref{g117}) --
(\ref{g118}). 
\end{prop}

Let
the common symbol $D$ stand for the operators $d_H$, $\dl$ and
$\vr$  in the complexes
(\ref{g117}) -- (\ref{g118}), and let $\G^*_\infty$ denote the
terms of these complexes. 
Since cohomology groups of these complexes  
are either trivial or
equal to the de Rham cohomology of $X$, 
 one can say that any
$D$-closed element
$\f\in \G^*_\infty$ takes the form 
\mar{g230}\beq
\f=\psi + D\xi,\qquad \xi\in \G^*_\infty,  \label{g230}
\eeq
where $\psi$
is a closed exterior form on $X$ which is not necessarily exact.
Since all $D$-closed elements of $\G^*_\infty$ of finite jet
order are also of form (\ref{g230}), it suffices to show that,
if an element
$\f\in
\cS^*_\infty$ is
$D$-exact in the module $\G^*_\infty$ (i.e., $\f=D\xi$, $\xi\in
\G^*_\infty$), then it is also in $\cS^*_\infty$ (i.e., 
$\f=D\vf$, $\vf\in \cS^*_\infty$).

Let $X$ be a (contractible) domain, and let an element 
$\f\in\cS^*_\infty$ be $D$-exact in $\G^*_\infty$. Then, being
$D$-closed, it is $D$-exact in $\cS^*_\infty$ in
accordance with Lemmas \ref{0465} and \ref{g220}.
Moreover,
a glance at the expressions (\ref{0473}), (\ref{0474}) and
(\ref{0477'}) shows that the maximal jet order
$[\vf]$ of $\vf$ is bounded by an integer $N([\f])$
which depends only on the maximal jet order
$[\f]$ of
$\f$. It follows that, if $\f=D\vf$ is an arbitrary 
$D$-exact form of the
jet order less than $k$, then the jet order of $\vf$ does not
exceed $N(k)$.   We agree to call
this fact the finite exactness of an operator
$D$. 

Let $X$ be an arbitrary manifold and $U$ a domain of $X$. By
virtue of Lemmas
\ref{0465} and \ref{g220}, 
the restriction of
the operator $D$ to 
$\G^*_\infty|_U$ (or, roughly speaking, the operator $D$ on
$U$) has the finite exactness property. Let us
state the following.

\begin{lem} \label{g222} \mar{g222}
Given a family $\{U_\al\}$ of disjoint open subsets of $X$, let
us suppose that the finite exactness of the operator
$D$ takes place on each subset $U_\al$ separately. Then $D$ 
on the union
$\op\cup_\al U_\al$ also has the finite exactness property.
\end{lem}

\begin{proof} Let
$\f\in\cS^*_\infty$ be a $D$-exact graded form on
$X$.
The finite exactness on $\cup
U_\al$ holds since $\f=D\varphi_\al$ on every
$U_\al$ and all $[\varphi_\al]<N([\f])$. 
\end{proof}

\begin{lem} \label{g223} \mar{g223}
Suppose that the finite exactness of an operator $D$ takes
place on open subsets
$U$, $V$ of $X$ and their non-empty overlap $U\cap V$. Then it
is also true on $U\cup V$.
\end{lem}

\begin{proof} Let
$\f=D\varphi\in\cS^*_\infty$ be a $D$-exact graded form on
$X$. By assumption, it can be brought into the form
$D\varphi_U$ on $U$ and $D\varphi_V$ on
$V$, where
$\varphi_U$ and $\varphi_V$ are graded forms of bounded
jet order.  Due to the decomposition (\ref{g230}), one can
choose the forms $\vf_U$, $\vf_V$ such that $\vf-\vf_U$ on $U$
and
$\vf-\vf_V$ on $V$ are $D$-exact. 
Let us consider the difference
$\varphi_U-\varphi_V$ on 
$U\cap V$.
It is a $D$-exact graded form of bounded
jet order which, by assumption, can be written as 
$\varphi_U-\varphi_V=D\si$ where 
$\si$ is also of bounded jet order. 
Lemma
\ref{am20} below shows that $\si=\si_U +\si_V$ where
$\si_U$ and
$\si_V$ are graded forms of bounded jet order on
$U$ and
$V$, respectively. Then, putting
\be
\varphi'_U=\varphi_U-D\si_U, \qquad  
\varphi'_V=\varphi_V+ D\si_V,
\ee
we have the graded form $\f$, equal to $D\varphi'_U$ on $U$ and 
$D\varphi'_V$ on $V$, respectively. Since the difference
$\vf'_U-\vf'_V$ on $U\cap V$ vanishes, we obtain $\f=D\vf'$ on
$U\cup V$  where 
\be
\vf'\op=^{\rm def}\left\{
\begin{array}{ccc}
\vf'|_U &= & \vf'_U\\
\vf'|_V &= & \vf'_V
\end{array}\right.
\ee
is of bounded jet order. 
\end{proof}

\begin{lem} \label{am20} \mar{am20}
Let $U$ and $V$ be open subsets of $X$ and $\si$
 a graded form of bounded jet order on
$U\cap V$. Then $\si$
splits into the sum $\si_U+ \si_V$ of graded exterior
forms $\si_U$ on
$U$ and
$\si_V$ on $V$ of bounded jet order. 
\end{lem} 

\begin{proof}
By taking a smooth partition of unity on $U\cup V$ subordinate 
to its cover
$\{U,V\}$ and passing to the function with support in $V$, 
we get a
smooth real function
$f$ on
$U\cup V$ which is 0 on a neighborhood $U_{U-V}$ of $U-V$ and 1
on a neighborhood $U_{V-U}$ of
$V-U$ in $U\cup V$. The graded form
$f\si$ vanishes on $U_{U-V}\cap (U\cap V)$ and, therefore, can
be extended by 0 to $U$. Let us denote it $\si_U$.
Accordingly, the graded form
$(1-f)\si$ has an extension $\si_V$ by 0 to 
$V$. Then $\si=\si_U +\si_V$ is a desired
decomposition because $\si_U$ and $\si_V$
are of finite jet order which does not exceed that of $\si$.
\end{proof}

Lemma 9.5 in \cite{bred}, Chapter V, states that, if some
property holds on a domain and obeys the conditions of Lemma
\ref{g222} and
\ref{g223}, it holds on any open subset of $\Bbb
R^n$. Hence, the operator $D$ has the jet exactness property on
any open subset of $\Bbb R^n$ and, consequently, on any chart of
the fiber bundle $Q\times_X Y\to X$. Since the latter admits a
finite  bundle atlas with the
transition functions (\ref{jet1}) and (\ref{+471}) preserving
the jet order, the finite exactness of $D$ takes place on the
whole manifold $X$ in accordance with Lemma \ref{g223}. This
proves Proposition \ref{g239} and, consequently, Theorem
\ref{g96}.

 \begin{rem} Let us consider the complex
\mar{g110}\beq
0\to\Bbb R\ar \cS^0_\infty\ar^d \cS^1_\infty\cdots
\ar^d\cS^k_\infty \ar\cdots, \label{g110}
\eeq
which we agree to call the de Rham complex because
$(\cS^*_\infty,d)$ is the differential calculus over the $\Bbb
R$-ring $\cS^0_\infty$. If $X=\Bbb R^n$, it is exact 
(\cite{drag}, Theorem 3.1). Similarly to the proof of Theorem
\ref{g96}, one can show that the cohomology of the de Rham
complex (\ref{g110}) equals the de Rham cohomology of $X$.
\end{rem}

\begin{cor} \label{cmp26} \mar{cmp26}
Every $d_H$-closed graded form $\f\in\cS^{0,m<n}_\infty$
falls into the sum
\mar{g214}\beq
\f=\psi + d_H\xi, \qquad \xi\in \cS^{0,m-1}_\infty, \label{g214}
\eeq 
where $\psi$ is a closed $m$-form on $X$. Every
$\dl$-closed graded Lagrangian $L\in \cS^{0,n}_\infty$ is the
sum
\mar{g215}\beq
\f=\psi + d_H\xi, \qquad \xi\in \cS^{0,n-1}_\infty, \label{g215}
\eeq
where $\psi$ is a non-exact $n$-form on $X$. 
\end{cor}

The global exactness of the complex (\ref{g112}) at the term
$\cS^{1,n}_\infty$ results in the following.

\begin{prop} \label{g103} \mar{g103}
Given a graded Lagrangian $L=\cL\om$, there
is the decomposition 
\mar{g99,'}\ben
&& dL=\dl L - d_H\Xi,
\qquad \Xi\in \cS^{1,n-1}_\infty, \label{g99}\\
&& \Xi=\op\sum_{s=0}
\th^A_{\nu_s\ldots\nu_1}\w
F^{\la\nu_s\ldots\nu_1}_A\om_\la,\qquad 
F_A^{\nu_k\ldots\nu_1}=
\dr_A^{\nu_k\ldots\nu_1}\cL-d_\la F_A^{\la\nu_k\ldots\nu_1}
+h_A^{\nu_k\ldots\nu_1},  \label{g99'}
\een
where local graded functions $h$ obey the relations
$h^\nu_a=0$,
$h_a^{(\nu_k\nu_{k-1})\ldots\nu_1}=0$. 
\end{prop}

\begin{proof} The decomposition (\ref{g99}) is a
straightforward consequence of the exactness of the
complex (\ref{g112}) at the term
$\cS^{1,n}_\infty$ and the fact that $\vr$ is a projector.
The coordinate expression (\ref{g99'}) results from
a direct computation
\be
&& -d_H\Xi= -d_H[\th^A F^\la_A+\th^A_\nu F^{\la\nu}_A +\cdots
+\th^A_{\nu_s\ldots\nu_1}
F^{\la\nu_s\ldots\nu_1}_A +\th^A_{\nu_{s+1}\nu_s\ldots\nu_1}\w
F^{\la\nu_{s+1}\nu_s\ldots\nu_1}_A+ \cdots]\w\om_\la=\\
&& \qquad [\th^Ad_\la F^\la_A +\th^A_\nu (F^\nu_A +d_\la
F^{\la\nu}_A)+\cdots
+\th^A_{\nu_{s+1}\nu_s\ldots\nu_1}
(F^{\nu_{s+1}\nu_s\ldots\nu_1}_A +
d_\la F^{\la\nu_{s+1}\nu_s\ldots\nu_1}_A) +\cdots]\w\om=\\
&& \qquad [\th^Ad_\la F^\la_A +\th^A_\nu (\dr^\nu_A\cL)+\cdots
+
\th^A_{\nu_{s+1}\nu_s\ldots\nu_1}
(\dr^{\nu_{s+1}\nu_s\ldots\nu_1}_A\cL)+\cdots]\w\om=\\
&& \qquad \th^A
(d_\la F^\la_A-\dr_A\cL) \w\om + dL= -\dl L+dL.
\ee
\end{proof}

Proposition \ref{g103} states the existence of a global
finite order Lepagean equivalent 
$\Xi_L=\Xi+L$ of any graded Lagrangian $L$. Locally, one can
always choose $\Xi$ (\ref{g99'}) where all functions $h$ vanish.

\section{Contact supersymmetries}

A graded derivation  $\up\in\gd
\cS^0_\infty$ of the $\Bbb R$-ring $\cS^0_\infty$ is said to
be an infinitesimal contact supertransformation or, simply, a
contact supersymmetry if the Lie derivative
$\bL_\up$ preserves the ideal of contact graded forms of the
BGDA
$\cS^*_\infty$ (i.e., the Lie derivative $\bL_\up$ of a
graded contact form is a graded contact form).  

\begin{prop} \label{g231} \mar{g231} With respect to the local
basis $(x^\la,s^A_\La, dx^\la,\th^A_\La)$ for the BGDA
$\cS^*_\infty$, any contact supersymmetry takes the form
\mar{g105}\beq
\up=\up_H+\up_V=\up^\la d_\la + (\up^A\dr_A +\op\sum_{|\La|>0}
d_\La\up^A\dr_A^\La), \label{g105}
\eeq
where $\up^\la$, $\up^A$ are local graded functions.
\end{prop}

\begin{proof}
The key point is that, since elements of $\cC^*_\infty$ can
be identified  as sections of a finite-dimensional vector bundle
over $X$, so can elements of the $C^\infty(X)$-algebra
$\cS^*_\infty$. Moreover, any graded form is a finite
composition of
$df$,
$f\in\cS^0_\infty$. Therefore, the proof follows that of
Proposition \ref{g62}.
\end{proof}

The interior product $\up\rfloor\f$ 
and the Lie derivative $\bL_\up\f$,
$\f\in\cS^*_\infty$ are defined   
by the same formulae
\be
&& \up\rfloor \f=\up^\la\f_\la + (-1)^{[\f_A]}\up^A\f_A, \qquad
\f\in \cS^1_\infty,\\ 
&& \up\rfloor(\f\w\si)=(\up\rfloor \f)\w\si
+(-1)^{|\f|+[\f][\up]}\f\w(\up\rfloor\si), \qquad \f,\si\in
\cS^*_\infty \\
&& \bL_\up\f=\up\rfloor d\f+ d(\up\rfloor\f), \qquad
\bL_\up(\f\w\si)=\bL_\up(\f)\w\si
+(-1)^{[\up][\f]}\f\w\bL_\up(\si).
\ee
as those on a graded manifold.
Following the proof of Lemma \ref{0410}, one can 
justify that any vertical contact supersymmetry
$\up$ (\ref{g105}) satisfies the relations 
\mar{g232,3}\ben
&& \up\rfloor d_H\f=-d_H(\up\rfloor\f), \label{g232}\\
&& \bL_\up(d_H\f)=d_H(\bL_\up\f), \qquad \f\in\cS^*_\infty.
\label{g233}
\een 

\begin{prop}  \label{g106} \mar{g106}
Given a graded Lagrangian $L\in\cS^{0,n}_\infty$, its 
Lie derivative $\bL_\up L$
along a contact
supersymmetry
$\up$ (\ref{g105}) fulfills the first variational formula
\mar{g107}\beq
\bL_\up L= \up_V\rfloor\dl L +d_H(h_0(\up\rfloor \Xi_L)) 
+ d_V (\up_H\rfloor\om)\cL, \label{g107}
\eeq
where $\Xi_L=\Xi+L$ is a Lepagean equivalent of $L$ given by
the coordinate expression (\ref{g99'}).
\end{prop}

\begin{proof} The proof follows that 
 of Proposition \ref{g75} and results from the decomposition
(\ref{g99}) and the relation (\ref{g232}). 
\end{proof}

In particular, let $\up$ be a variational
symmetry of a graded Lagrangian $L$, i.e., $\bL_\up
L=d_H\si$,
$\si\in
\cS^{0,n-1}_\infty$. 
Then the first variational formula (\ref{g107}) 
restricted to Ker$\,\dl L$ leads to the weak
conservation law
\mar{g108}\beq
0\ap d_H(h_0(\up\rfloor\Xi_L)-\si). \label{g108}
\eeq

\begin{rem} \label{g234} \mar{g234}
Let us consider the gauge theory of principal connections on
a principal bundle
$P\to X$ with a structure Lie group $G$. These
connections are represented by sections of the quotient 
$C=J^1P/G\to X$ \cite{book00}. This is an affine
bundle coordinated by
$(x^\la, a^r_\la)$ such that, given a section $A$ of $C\to X$, 
its components $A^r_\la=a^r_\la\circ A$ are coefficients of
the familiar local connection form (i.e., gauge potentials).
Let $J^\infty C$ be the infinite order jet
manifold of $C\to X$ coordinated by
$(x^\la,a^r_{\la,\La})$, $0\leq |\La|$, and let
$\cP^*_\infty(C)$ be the polynomial subalgebra of the GDA
$\cO^*_\infty(C)$. Infinitesimal generators of one-parameter
groups of vertical automorphisms  (gauge transformations) of a
principal bundle $P$ are
$G$-invariant vertical vector fields on $P\to X$.
They are associated
to sections of the vector bundle
$V_GP=VP/G\to X$ of right Lie algebras of the group $G$. Let us
consider the simple graded manifold
$(X,\gA_{V_GY})$ constructed from this vector bundle. Its local
basis is $(x^\la, C^r)$. Let $\cC^*_{J^rV_GY}$ be the  
BGDA of graded
exterior forms on the graded manifold $(X,\gA_{J^rV_GP})$, and
$\cC^*_\infty(V_GP)$ the direct limit of the direct system
(\ref{g205}) of these algebras. Then the graded product
\mar{g211}\beq
\cS^*_\infty(V_G,C)=\cC^*_\infty(V_GP)\w \cP^*_\infty(C)
\label{g211}
\eeq
describes gauge potentials, odd ghosts and their jets in the
BRST theory. With respect
to a local basis
$(x^\la,a^r_\la, C^r)$ for the BGDA $\cS^*_\infty(V_G,C)$
(\ref{g211}),  the BRST symmetry is given by the
contact supersymmetry
\mar{g130}\ben
&& \up= \up_\la^r\dr^\la_r +
\up^r\dr_r
+\op\sum_{|\La|>0}(d_\La\up_\la^r\dr^{\La,\la}_r +
d_\La\up^r\dr_r^\La),
\label{g130}\\ 
&& \up_\la^r=C_\la^r +c^r_{pq}a^p_\la C^q,
\qquad \up^r=-
\frac12c^r_{pq}C^p C^q, \nonumber 
\een
where $c^r_{pq}$ are structure constants of the Lie
algebra of $G$ and $\dr^\la_r$, $\dr_r$, $\dr^{\La,\la}_r$
$\dr_r^\La$ are the duals of  
$da^r_\la$,
$dC^r$, $da^r_{\La,\la}$ and $dC^r_\La$, respectively. A
remarkable peculiarity of this contact supersymmetry is that
the Lie derivative
$\bL_\up$ along $\up$ (\ref{g130}) is nilpotent on the module
$S^{0,*}_\infty$ of horizontal graded forms.
\end{rem}

In a general setting, a vertical contact
supersymmetry
$\up$ (\ref{g105}) is said to be nilpotent if 
\mar{g133}\beq
\bL_\up(\bL_\up\f)= \op\sum_{|\Si|\geq 0,|\La|\geq 0 }
(\up^B_\Si\dr^\Si_B(\up^A_\La)\dr^\La_A + 
(-1)^{[s^B][\up^A]}\up^B_\Si\up^A_\La\dr^\Si_B \dr^\La_A)\f=0
\label{g133}
\eeq
for any horizontal graded form $\f\in S^{0,*}_\infty$. 

\begin{lem} \label{041} \mar{041} A contact supersymmetry
$\up$ is  nilpotent iff it is odd and the equality
\be
\bL_\up(\up^A)=\op\sum_{|\Si|\geq 0} \up^B_\Si\dr^\Si_B(\up^A)=0
\ee
holds for all $\up^A$.
\end{lem}

\begin{proof} There is the relation 
\mar{0490}\beq
d_\la\circ v^i_\La\dr_i^\la= v^i_\La\dr_i^\la \circ d_\la,
\label{0490}
\eeq
similar to (\ref{0480'}). Then the lemma follows from the
equality (\ref{g133}) where one puts $\f=s^A$ and $\f=s^A_\La
s^B_\Si$.
\end{proof}

\begin{rem} \label{g235} \mar{g235}
A useful example of a nilpotent contact
supersymmetry is the supersymmetry
\mar{g134}\beq
\up=\up^A(x)\dr_A +\op\sum_{|\La|>0}\dr_\La \up^A\dr_A^\La,
\label{g134} 
\eeq
where all $\up^A$ are smooth real functions on $X$, $\dr_A^\la$
are the duals of $ds^A_\La$, but all
$s^A$ are odd.
\end{rem}

\section{Cohomology of nilpotent contact supersymmetries}

Let $\up$ be a nilpotent contact supersymmetry.
Since the Lie derivative $\bL_\up$ obeys the relation
(\ref{g233}), let us assume that the $\Bbb
R$-module $\cS^{0,*}_\infty$ of graded horizontal forms is
split into a bicomplex $\{S^{k,m}\}$ with respect
to the nilpotent operator $\bs_\up$ (\ref{g240}) and the total
differential
$d_H$ which obey the relation 
\mar{04101}\beq
\bs_\up\circ d_H+ d_H\circ
\bs_\up=0. \label{04101}
\eeq
This bicomplex 
\be
d_H: S^{k,m}\to S^{k,m+1}, \qquad \bs_\up: S^{k,m}\to
S^{k+1,m} 
\ee
is graded by the form degree $0\leq m\leq
n$ and an integer $k\in \Bbb Z$, though it may happen that
$S^{k,*}=0$ starting from some number $k$. For the sake
of brevity, let us call
$k$ the charge number. 

For instance, the BRST bicomplex
$S^{0,*}_\infty(C,V_GP)$ is graded by the charge number $k$
which is the polynomial degree of its elements 
 in odd variables $C^r_\La$. In this case, $\bs_\up$
(\ref{g240}) is the BRST operator. Since the ghosts $C^r_\La$
are characterized by the ghost number 1, $k\in \Bbb N$, is the
ghost number. The bicomplex defined by the contact supersymmetry
(\ref{g134})  has the similar gradation, but taken with the sign
minus (i.e., $k=0,-1,\ldots$) because the nilpotent
operator $\bs_\up$ decreases the odd polynomial
degree. 

Let us consider the relative and iterated cohomology
of the nilpotent operator
$\bs_\up$ (\ref{g240}) with respect to the total differential
$d_H$.
Recall that a horizontal graded form $\f\in S^{*,*}$ is said to
be  a relative closed form, i.e., $(\bs_\up/d_H)$-closed form
if
$\bs_\up\f$ is a
$d_H$-exact form. This form is called
exact if it is a sum of an $\bs_\up$-exact form and a
$d_H$-exact form. Accordingly, we have the
relative cohomology $H^{*,*}(\bs_\up/d_H)$. If a
$(\bs_\up/d_H)$-closed form $\f$ is also $d_H$-closed, it is
called an iterated 
$(\bs_\up|d_H)$-closed
form. This form $\f$ is said to be exact if $\f=\bs_\up\xi
+d_H\si$, where $\xi$ is a $d_H$-closed form. Thus, we obtain
the iterated cohomology 
$H^{*,*}(\bs_\up|d_H)$ of the
$(\bs_\up,d_H)$-bicomplex
$S^{*,*}$. It is the term $E_2^{*,*}$ of the
spectral sequence of this bicomplex
\cite{mcl}. There is an obvious isomorphism 
$H^{*,n}(\bs_\up/d_H)=H^{*,n}(\bs_\up|d_H)$ of relative  and
iterated cohomology groups on horizontal graded densities. 
Forthcoming Theorems \ref{g250} and
\ref{g251} extend our results on iterated cohomology in
\cite{lmp} to an arbitrary nilpotent contact supersymmetry. 

\begin{prop} \label{g241} \mar{g241}
Let us consider the complex
\mar{g238}\beq
0\ar \Bbb R\ar
\cS^0_\infty\ar^{d_H}\cS^{0,1}_\infty \cdots
\ar^{d_H} \cS^{0,n}_\infty\ar^{d_H} 0.
\label{g238} 
\eeq 
Its cohomology groups $H^{m<n}(d_H)$ 
equal  the de Rham cohomology groups $H^{m<n}(X)$ of $X$, while
the cohomology group $H^n(d_H)$ fulfills the
relation
\mar{g242}\beq
H^n(d_H)/H^n(X)=E_1. \label{g242}
\eeq
\end{prop}

\begin{proof} The complex (\ref{g238}) differs from
the short variational complex (\ref{g111}) in the last term.
Therefore its cohomology $H^{m<n}(d_H)$ equals the cohomology
of the complex (\ref{g111}) of the form degree $m<n$. The
formula (\ref{g242}) follows from the 
relations:
(i) $H^n(d_H)=\cS^{0,n}_\infty/d_H(\cS^{0,n-1}_\infty)$, (ii) 
$E_1=\cS^{0,n}_\infty/\Ker \dl$, since
$\dl$ in the complex (\ref{g111}) is an epimorphism, and 
(iii) $\Ker \dl/d_H(\cS^{0,n-1}_\infty)=H^n(X)$ owing to the 
formula (\ref{g215}).
\end{proof}

\begin{theo} \label{g250} \mar{g250}
There is an epimorphism
\mar{g252}\beq
\zeta:
H^{m<n}(X)\to H^{*,m<n}(\bs_\up|d_H) \label{g252}
\eeq 
of the de Rham cohomology $H^m(X)$ of $X$ 
of form degree less than
$n$ onto the iterated
cohomology $H^{*,m<n}(\bs_\up|d_H)$.
\end{theo}

\begin{proof}
Since a nilpotent
contact supersymmetry $\up$ is vertical, all
exterior forms
$\f$ on $X$ are $\bs_\up$-closed. It follows that $d$-cocycles
on $X$ are
$(\bs_\up|d_H)$-closed. Since any $d_H$-exact horizontal graded
form is also $(\bs_\up|d_H)$-exact, we have     
a morphism $\zeta$ (\ref{g252}). By virtue of
Corollary
\ref{cmp26} (and, equivalently, Proposition \ref{g241}), any
$d_H$-closed horizontal graded
$(m<n)$-form $\f$ is split into the sum $\f=\vf
+d_H\xi$ (\ref{g214}) of a closed $m$-form $\vf$ on $X$ and a
$d_H$-exact graded form. Therefore, any 
$(\bs_\up|d_H)$-cocycle is the sum of a closed exterior form on
$X$ and a $d_H$-exact graded form. 
It follows that the morphism $\zeta$
(\ref{g252}) is an epimorphism.  
The kernel of the morphism $\zeta$ (\ref{g252}) consists of
elements whose representatives are $\bs_\up$-exact
closed exterior forms on $X$.
\end{proof}

In particular, if $X=\Bbb R^n$, the iterated cohomology
$H^{*,0<m<n}(\bs_\up|d_H)$ is trivial in contrast with the
relative ones.

 For instance, 
since coefficients of the BRST
transformation $\up$ (\ref{g130}) 
consist of polynomials in ghosts $C^r$ of non-zero degree, 
exterior forms on
$X$ are never
$\bs_\up$-exact and, consequently, $\zeta$ (\ref{g252})  is an
isomorphism. However, this is not the case of the supersymmetry
(\ref{g134}).  

\begin{cor} \label{0496} \mar{0496}
If exterior forms on $X$ are only of zero charge number, 
the iterated cohomology 
$H^{\neq 0,m<n}(\bs_\up|d_H)$ is trivial and $\zeta$
(\ref{g252}) is an epimorphism 
$H^{m<n}(X) \to
H^{0,m<n}(\bs_\up|d_H)$.
\end{cor}

In particular, this is the case of the contact supersymmetries
(\ref{g130}) and (\ref{g134}). Moreover, if $\up$ is
the BRST transformation (\ref{g130}), $\zeta$ in Corollary
\ref{0496} is an isomorphism.

The bicomplex $S^{*,*}$ is
a complex with respect to the total coboundary operator
$\wt\bs_\up=\bs_\up+ d_H$.
 We intend to determine the relation between the iterated
cohomology
$H^{*,m}(\bs_\up|d_H)$ and the total
$\wt\bs_\up$-cohomology $H^*(\wt\bs_\up)$ of the bicomplex
$S^{*,*}$. 

There exists the morphism
\mar{g255}\beq
\g: H^{<n}(X)\to H^*(\wt\bs_\up) \label{g255}
\eeq
of the de Rham cohomology $H^{<n}(X)$ of $X$ 
of form degree 
$<n$ to the total
cohomology $H^*(\wt\bs_\up)$ 
similar to the 
morphism (\ref{g252}). The morphism $\g$ (\ref{g255})
associates to a closed $m$-form $\f$ of charge number $k$ the
$\wt s_\up$-cocycle of total charge $(k+m)$ whose
representative is $\f$. Its kernel consists of elements whose
representatives are
$\wt\bs_\up$-exact closed exterior forms on $X$. 

\begin{theo} \label{04100} \mar{04100} 
There is a monomorphism of the iterated cohomology
$H^{*,m<n}(\bs_\up|d_H)$ to the total cohomology
$H^*(\wt\bs_\up)$ which associates to an iterated
$(k,m)$-cocycle $[\f]$ 
the $\wt s_\up$-cocycle of total charge $(k+m)$
represented by the same graded form $\f$. 
\end{theo}

\begin{proof} Any $(\bs_\up|d_H)$-cocycle, by definition, is a
$\wt s_\up$-cocycle. Using the formula (\ref{g214}) and the
fact that exterior forms on $X$ are $s_\up$-closed, one can show
that: (i) any 
$(\bs_\up|d_H)$-coboundary is also a $d_H$-coboundary and,
consequently, a $\wt s_\up$-coboundary, and (ii) if a
$(\bs_\up|d_H)$-cocycle is a $\wt s_\up$-coboundary, then it is
a $(\bs_\up|d_H)$-coboundary. This proves the statement.
\end{proof}

Turn now to the iterated cohomology $H^{*,n}(\bs_\up|d_H)$.
This requires careful analysis since Proposition \ref{g241}
implies that the cohomology
$H^n(d_H)$ of the complex (\ref{g238}) fails to equal the de
Rham cohomology
$H^n(X)$ of $X$. 

\begin{theo} \label{g251} \mar{g251}
Put $\ol H^*=
H^*(\wt\bs_\up)/\im \g$, where the asterisk means the total
charge. There is an
isomorphism 
\mar{g270}\beq
H^{*,n}(\bs_\up|d_H)/\ol H^*=\Ker \g. \label{g270}
\eeq
\end{theo}

\begin{proof} The proof falls into the following three steps.

(i) First, we show the existence of a morphism
\mar{g271}\beq
\eta:H^{*,n}(\bs_\up|d_H)\to \Ker \g \label{g271}
\eeq
from the iterated cohomology group $H^{*,n}(\bs_\up|d_H)$ to
$\Ker\g$. Consider a horizontal graded $n$-form $\f_n$ which is
$(\bs_\up|d_H)$-closed. Then, by definition,  
$\bs_\up\f_n$ is
$d_H$-exact, i.e.,
\mar{aa10}\beq
\bs_\up\f_n + d_H\f_{n-1}=0. \label{aa10}
\eeq
Acting on this equality by $\bs_\up$, we
observe that
$\bs_\up\f_{n-1}$ is a 
$d_H$-closed graded form, i.e.,
\mar{g140}\beq
\bs_\up\f_{n-1} + d_H\f_{n-2}=\vf_{n-1}, \label{g140}
\eeq
where $\vf_{n-1}$ is a closed $(n-1)$-form on $X$ in accordance
with Corollary
\ref{cmp26}. Since
$\bs_\up\vf_{n-1}=0$, an action of $\bs_\up$ on the equation
(\ref{g140}) shows that 
$\bs_\up\f_{n-2}$ is a 
$d_H$-closed graded form, i.e.,
\be
\bs_\up\f_{n-2} + d_H\f_{n-3}=\vf_{n-2}, 
\ee
where $\vf_{n-2}$ is a closed $(n-2)$-form on $X$.
Iterating the
arguments, one comes to the system of equations
\mar{aa11}\beq
\bs_\up\f_{n-k} + d_H\f_{n-k-1}=\vf_{n-k},
\qquad 0\leq k <n, \qquad
\bs_\up \f_0=\vf_0={\rm const},\label{aa11}
\eeq
which can be assembled into descent equations 
\mar{g256,aa13}\ben
&& \wt\bs_\up\wt \f=\wt\vf, \label{g256} \\
&& \wt \f=\f_n+\f_{n-1}+\cdots +\f_0, \qquad \wt\vf=
\vf_{n-1}+\cdots +\vf_0.
\label{aa13}
\een
Thus, any $(\bs_\up|d_H)$-closed  horizontal graded form defines
descent equations (\ref{g256}) whose right-hand sides
$\wt\vf$ are closed exterior forms on $X$ such that their
de Rham classes belong to the kernel
$\Ker\g$ of the morphism (\ref{g255}). 
For the sake
of brevity, let us denote these descent equations by
$\lng\wt\vf\rng$. Accordingly, we say that a
horizontal  graded form 
$\wt \f$ (\ref{aa13}) is a solution of descent equations
$\lng\wt\vf\rng$ (\ref{aa13}). Descent equations defined
by a $(\bs_\up|d_H)$-closed horizontal graded form $\f_n$ are
not unique. Let $\wt\f'$ be another solution of another
set of descent equations $\lng\wt\vf'\rng$
such that $\f_n=\f'_n$. Let us denote
$\Delta\f_k=\f_k-\f'_k$ and 
$\Delta\vf_k=\vf_k-\vf'_k$. Then the equations (\ref{aa10})
lead  to the equation $d_H(\Delta\f_{n-1})=0$. It follows
from Corollary \ref{cmp26} that
\mar{g257}\beq
\Delta\f_{n-1}=d_H\xi_{n-2} +\al_{n-1}, \label{g257}
\eeq
where $\al_{n-1}$ is a closed $(n-1)$-form on $X$. Accordingly,
the equation (\ref{aa11}) leads to the equation
\be
\bs_\up(\Delta\f_{n-1})+d_H(\Delta\f_{n-2})=\Delta\vf_{n-1}.
\ee
Substituting the
equality (\ref{g257}) into this equation and bearing in mind
the relation (\ref{04101}), we obtain the equality
\be
d_H(-\bs_\up\xi_{n-2}+ \Delta\f_{n-2})= \Delta\vf_{n-1}.
\ee
It follows that 
\be
\Delta\f_{n-2}
=\bs_\up\xi_{n-2} + d_H\xi_{n-3} +\al_{n-2}, \qquad
\Delta\vf_{n-1}=d\al_{n-2}
\ee
where $\al_{n-2}$ is an exterior form on $X$. Iterating the
arguments, one comes to the relations
\mar{g258}\beq
\Delta\f_{n-k} =\bs_\up\xi_{n-k} + d_H\xi_{n-k-1} + \al_{n-k},
\qquad
\Delta\vf_{n-k}=d \al_{n-k-1}, \qquad 1<k<n,
\label{g258}
\eeq
where $\al_{n-k-1}$ are exterior forms on $X$ and, finally, to
the equalities $\Delta\f_0=0$, $\Delta\vf_0=0$.
Then it
is easily justified that 
\mar{g259,'}\ben
&& \wt\f-\wt\f'=\wt\bs_\up\wt\si +\wt\al, 
\qquad \wt\si= \xi_{n-2}+\cdots \xi_1,
\label{g259}\\ && \wt\vf - \wt\vf'=d\wt\al, \qquad
\wt\al= \al_{n-1}+\cdots +\al_1. \label{g259'}
\een
It follows that right-hand sides of any two descent equations
defined by a $(\bs_\up|d_H)$-closed horizontal graded form
$\f_n$ differ from each other in an exact form on $X$. 
Moreover,  let $\f_n$ and $\f'_n$ be representatives of 
the same iterated cohomology class in $H^{*,n}(\bs_\up|d_H)$,
i.e., $\f'_n=\f_n+ \bs_\up\psi +d_H\bt$, where $\psi$ is
$d_H$-closed. Let
$\f_n$ provide a solution $\wt\f$ of a descent equation 
$\lng\wt\vf\rng$. Then $\f'_n$ defines a 
solution 
$\wt\f'=\wt\f+\wt\bs_\up(\psi+\bt)$ of the same descent
equation.
Thus, the assignment 
$\f_n\mapsto \lng\wt \vf\rng$ yields the desired
morphism $\eta$ (\ref{g271}).

(ii) Let us show that the morphism $\eta$ (\ref{g271}) is an
epimorphism. Let $\wt\vf$ be a
closed exterior form  on
$X$ whose de Rham cohomology class belongs to
$\Ker
\g$. Let $\wt\vf=
\vf_{n-1}+\cdots +\vf_0$ be its decomposition in $k$-forms
$\vf_k$,
$k=1,\ldots,n-1$. Then the family of exterior
forms
$(\vf_k)$  yields a system of the equations (\ref{aa11}) which
can be assembled into the descent equations
$\lng\wt\vf\rng$ (\ref{g256}). Its solution $\wt \f$
exists because $\wt\vf\in\Ker \g$. Let
$\wt\vf'$ differ from 
$\wt\vf$ in an exact form, i.e., let the relation
(\ref{g259'}) hold. Then any solution
$\wt
\f$ of the equation $\lng\wt\vf\rng$ yields a
solution
$\wt\f'=\wt\f-\wt\al$ (\ref{g259}) of the equation
$\lng\wt\vf'\rng$ such that $\f'_n=\f_n$.
It follows that the morphism $\eta$ (\ref{g271}) is an
epimorphism.

(iii) The kernel of the morphism $\eta$ (\ref{g271})
is represented by  $(\bs_\up|d_H)$-closed  horizontal
graded forms
$\f_n$ which yield  homogeneous descent equations 
\mar{g273}\beq
\wt\bs_\up\wt\f=0. \label{g273}
\eeq
Let us define an epimorphism of
the total cohomology $H^*(\wt\bs_\up)$ onto 
$\Ker\eta$. For this purpose, let us associate to each $\wt
s_\up$-cocycle $\wt\f$ its higher term $\f_n$. The latter
defines homogeneous descent equations (\ref{g273}) whose
solution is $\wt\f$, i.e., $\f_n\in\Ker\eta$. Let $\wt\f=\wt
s_\up\wt\psi$ be a $\wt s_\up$-coboundary. Its higher term
$\f_n$ takes the form $\f_n=s_\up\psi_n + d_H\psi_{n-1}$, i.e.,
it is an iterated coboundary. It follows that the assignment
 $\wt\f\mapsto\f_n$ provides the desired
epimorphism $\tau:H^*(\wt\bs_\up)\to \Ker\eta$. The kernel of
this epimorphism is represented by solutions $\wt\f$ of the
descent equation (\ref{g273}) whose higher term vanishes.
Following item (i), one can easily show that these solutions
take the form
$\wt\f=\wt\bs_\up\si+\wt\al$, where $\wt\al$ is a closed
exterior form on $X$ of form degree $<n$. Cohomology
classes of these solutions exhaust the image of the
morphism $\g$ (\ref{g255}), i.e., $\im\g=\Ker\tau$. 
\end{proof}

In particular, if the morphism $\g$ (\ref{g255}) is a
monomorphism (i.e., no non-exact closed exterior form on $X$ is 
$\wt\bs_\up$-exact), the isomorphism (\ref{g270}) gives the
isomorphism 
\be
H^{*,n}(\bs_\up|d_H)=H^*(\wt\bs_\up)/H^{<n}(X).
\ee
For instance, this is the case of the BRST transformation
(\ref{g130}) \cite{lmp}.

\section{Conclusion}

In the present work, 
we follow the algebraic topological approach to describing
Lagrangian field theories in terms of the variational
bicomplex. This enables us to extend the cohomology analysis of
Lagrangian BRST theory on $\Bbb R^n$ to a generic contact
supersymmetry on an arbitrary manifold $X$. Since only
vector and affine bundles over $X$ are involved, the
corresponding cohomology characteristic is represented by the
de Rham cohomology of
$X$. In a general case of nilpotent contact
supersymmetry $\bs_\up$,  its contribution however is
not  trivial because exterior forms on
$X$ are
$\bs_\up$-closed, but need not be $\bs_\up$-exact. For instance,
this contribution is given by the kernel of the morphism $\g$
(\ref{g255}) in Theorem \ref{g251}.
Our analysis  seems important for BV
quantization of field systems with non-contractible topologies,
e.g., gravitation theory and topological field models.
We also bear in mind the extension of BV quantization to
field systems where parameters of gauge transformations 
may depend on field variables and their derivatives
\cite{fulp02}. For instance, this is the case of spinor fields
in gauge gravitation theory \cite{sard98}.
\bigskip

\noindent
{\it Acknowledgement}. The authors would like to thank a referee
for carefully reading the manuscript and numerous suggestions.

\section{Appendixes}

\noindent
{\it Appendix A}. Since $Y$ is a strong deformation retract of
any finite order jet manifold $J^rY$, the de Rham cohomology of
the GDA
$\cO_\infty^*$  is easily proved to equal the de Rham
cohomology $H^*(Y)$ of $Y$ in accordance with Theorem
\ref{0460} \cite{ander}. However, we must enlarge  
$\cO_\infty^*$
in order to find its $d_H$- and $\dl$-cohomology.

\noindent 
{\it Outline of proof of Theorem \ref{g90}}. 
One starts from the algebraic Poincar\'e lemma
\cite{olv,tul}.

\begin{lem} \label{042} \mar{042} If $Y$ is a contractible
bundle $\Bbb R^{n+p}\to\Bbb R^n$, the variational bicomplex
(\ref{7}) is exact.
\end{lem}

For instance, the homotopy operators for $d_V$, $d_H$, $\dl$ and
$\vr$ are given by the formulae (5.72), (5.109), (5.84)
in \cite{olv} and (4.5) in \cite{tul}, respectively.

Let $\gO^*_r$ be the sheaf
of germs of exterior forms on the $r$-order jet 
manifold $J^rY$, and let
$\ol\gO^*_r$ be its canonical presheaf.
There is the direct  system
of presheaves
\be
\ol\gO^*_X\op\longrightarrow^{\pi^*} \ol\gO^*_0 
\op\longrightarrow^{\pi^1_0{}^*} \ol\gO_1^* \cdots
\op\longrightarrow^{\pi^r_{r-1}{}^*}
 \ol\gO_r^* \longrightarrow\cdots. 
\ee
Its direct limit $\ol\gO^*_\infty$ 
is a presheaf of GDAs on the infinite order jet manifold
$J^\infty Y$. Let
$\gQ^*_\infty$ be the sheaf of GDAs on $J^\infty Y$ constructed
from  the presheaf $\ol\gO^*_\infty$, i.e., $\gQ^*_\infty$  
is the shef of germs of $\ol\gO^*_\infty$ (we follow
the terminology of
\cite{hir}).  The structure module 
$\G(\gQ^*_\infty)$ of 
sections of $\gQ^*_\infty$ is a GDA such that, given an
element
$\f\in \G(\gQ^*_\infty)$ and a point $z\in J^\infty Y$, there 
exist an open
neighbourhood $U$ of $z$ and an
exterior form
$\f^{(k)}$ on some finite order jet manifold $J^kY$ so that
$\f|_U= \pi^{\infty*}_k\f^{(k)}|_U$. In
particular, there is the  monomorphism $\cO^*_\infty
\to\G(\gQ^*_\infty)$. The fact that 
the
paracompact space
$J^\infty Y$ admits a partition of unity by elements of
the ring $\G(\gQ^0_\infty)$ \cite{tak2},  enables one to obtain 
$d_H$- and $\dl$-cohomology of $\G(\gQ^*_\infty)$ as follows
\cite{and,ander,ijmms,tak2}. 

The sheaf $\gQ^*_\infty$ is split into the 
bicomplex $\gQ^{*,*}_\infty$. Let us consider its
variational subcomplex and the complexes of sheaves  of contact
forms
\mar{g91,'}\ben
&& 0\to\Bbb R\to \gQ^0_\infty
\ar^{d_H}\gQ^{0,1}_\infty\cdots  
\op\longrightarrow^{d_H} 
\gQ^{0,n}_\infty  \op\longrightarrow^\dl \gE_1 
\op\longrightarrow^\dl 
\gE_2 \longrightarrow \cdots, \qquad
\gE_k=\vr(\gQ^{k,n}_\infty), 
\label{g91} \\
&& 0\to \gQ^{k,0}_\infty\ar^{d_H} \gQ^{k,1}_\infty\cdots  
\ar^{d_H} 
\gQ^{k,n}_\infty  \ar^\vr
\gE_k\to 0,
\label{g91'} 
\een
together with complexes of their structure modules
\mar{g92,'}\ben
&& 0\to\Bbb R\to \G(\gQ^0_\infty)
\ar^{d_H}\G(\gQ^{0,1}_\infty)\cdots  
\op\longrightarrow^{d_H} 
\G(\gQ^{0,n}_\infty)  \op\longrightarrow^\dl \G(\gE_1) 
\op\longrightarrow^\dl 
\G(\gE_2) \longrightarrow \cdots, 
\label{g92} \\
&& 0\to \G(\gQ^{k,0}_\infty)\ar^{d_H}
\G(\gQ^{k,1}_\infty)\cdots  
\ar^{d_H} 
\G(\gQ^{k,n}_\infty)  \ar^\vr
\G(\gE_k)\to 0.
\label{g92'} 
\een
By virtue of Lemma
\ref{042} and Theorem \ref{0460}, the  complexes (\ref{g91})
-- (\ref{g91'}) are exact.
Since $\gQ^{*,*}_\infty$ are sheaves of 
$\G(\gQ^0_\infty)$-modules, they are fine.
 The sheaves $\gE_k$ are also proved to be fine
\cite{lmp,ijmms}. Consequently, all sheaves, except $\Bbb
R$, in the complexes (\ref{g91}) -- (\ref{g91'}) are acyclic.
Therefore, these complexes are 
the resolutions of the constant sheaf $\Bbb R$ 
and the zero sheaf over $J^\infty Y$, respectively. In
accordance with the abstract de Rham theorem (\cite{hir},
Theorem 2.12.1), cohomology of the complex (\ref{g92}) equals
the cohomology  of
$J^\infty Y$ with coefficients in
$\Bbb R$, while the complex (\ref{g92'}) is exact. Since 
$Y$ is a strong deformation retract of
$J^\infty Y$ \cite{ander,jmp}, cohomology  of the complex
(\ref{g92}) is isomorphic to  the de Rham cohomology of
$Y$. 

Note that, in order to prove the exactness of the complex
(\ref{g92'}), one can use a minor generalization of the above
mentioned abstract de Rham theorem (see Appendix D), and need
not   justify the acyclicity of the sheaves $\gE_k$
\cite{tak2}.

Finally, the
subalgebra 
$\cO^*_\infty\subset \G(\gQ^*_\infty)$ is proved to have
the same
$d_H$- and
$\dl$-cohomology as
$\G(\gQ^*_\infty)$ \cite{lmp,ijmms}. Similarly, one can show
that, restricted to $\cO^{k,n}_\infty$, the operator $\vr$
remains exact. $\Box$ \medskip

The following is a corollary of item (i) of Theorem
\ref{g90} (cf. Corollary \ref{cmp26}).

\begin{cor} \label{g212} \mar{g212}
Every $d_H$-closed form  $\f\in 
\cO^{0,m<n}$ is the sum 
\mar{g213}\beq
\f=h_0\psi +d_H\xi, \qquad \xi\in
\cO^{0,m-1}_\infty, \label{g213}
\eeq
where $\psi$ is a closed $m$-form on $Y$. 
Every $\dl$-closed Lagrangian
$L\in\cO^{0,n}_\infty$ is the sum
\mar{t42}\beq
 L=h_0\psi + d_H\xi,  \qquad \xi\in
\cO^{0,n-1}_\infty,
\label{t42}
\eeq
where $\psi$ is a closed $n$-form on $Y$.
\end{cor} 

Note that the formulae (\ref{g213}) -- (\ref{t42}) were obtained
in \cite{and} by computing cohomology of the fixed order
variational sequence, but the proof of the local exactness of
this sequence requires rather sophisticated {\it ad hoc}
techniques. 

\bigskip
\noindent
{\it Appendix B}.
Let us consider the open
surjection
$\pi^\infty: J^\infty Y\to X$ and the direct image 
$\pi^*_\infty\gQ^*_\infty$ on $X$ of the sheaf $\gQ^*_\infty$ 
of exterior forms on $J^\infty Y$. Its stalk at a point $x\in
X$ consists of the equivalence classes of sections of the sheaf
$\gQ^*_\infty$ which coincide on the inverse images
$(\pi^\infty)^{-1}(U_x)$ of open neighbourhoods $U_x$ of $X$.
Since $(\pi^\infty)^{-1}(U_x)$ is the infinite order jet
manifold of sections of the fiber bundle $\pi^{-1}(U_x)\to X$, 
every point
$x\in X$ has a base of open neighbourhoods
$\{U_x\}$ such that the sheaves $\gQ^{*,*}_\infty$ and 
$\gE_k$ in the proof of
Theorem \ref{g90} are acyclic on the inverse
images
$(\pi^\infty)^{-1}(U_x)$ of these neighbourhoods. Then, in
accordance with the Leray theorem \cite{god}, cohomology of
$J^\infty Y$ with coefficients in the sheaves $\gQ^{*,*}_\infty$
and $\gE_k$ is isomorphic to that of $X$ with coefficients in
their direct images $\pi^*_\infty\gQ^{*,*}_\infty$ and
$\pi^*_\infty\gE_k$, i.e., the sheaves 
$\pi^*_\infty\gQ^{*,*}_\infty$ and $\pi^*_\infty\gE_k$ over $X$
are acyclic. Let
$Y\to X$ be an affine bundle. Then $X$ is a strong
deformation retract of $J^\infty Y$. In this case, the  inverse
images 
$(\pi^\infty)^{-1}(U_x)$ of contractible neighbourhoods $U_x$
are contractible and $\pi^\infty_*\Bbb R=\Bbb R$. Then, by 
virtue of Lemma \ref{042}, the variational
bicomplex
$\gQ^*_\infty$ of sheaves over 
$(\pi^\infty)^{-1}(U_x)$ is exact,
and the variational bicomplex
$\pi^\infty_*\gQ^*_\infty$ of sheaves over $X$ is so. There is
an $\Bbb R$-algebra isomorphism of the GDA of sections of the
sheaf $\pi^\infty_*\gQ^*_\infty$ over $X$ to the GDA
$\G(\gQ^*_\infty)$. Thus, the GDA
$\G(\gQ^*_\infty)$ and its subalgebra
$\cO^*_\infty$ can be regarded as algebras  of sections of
a sheaf over $X$. 

\bigskip
\noindent
{\it Appendix C}.
Let us associate to each open subset $U\subset X$ the bigraded
algebra
$\cS^*_U$ of elements of the
$C^\infty(X)$-algebra $\cS^*_\infty$ whose coefficients
are restricted to
$U$. These algebras make up a presheaf 
\mar{0450}\beq
\{\cS^*_U, r^U_V \,\,|\,\, r^U_V: \cS^*_U\to \cS^*_V \}
\label{0450}
\eeq
over $X$. Let
$\gS^*_\infty$ be the sheaf constructed from this presheaf. Its
stalk $\gS^*_x$ at a point $x\in X$ is the direct limit of the
direct system of $\Bbb
R$-modules $\{\cS^*_U,r^U_V\}$, indexed by the directed set of
open neighbourhoods $U$ of $x$. This stalk consists of the germs
of elements of $\cS^*_\infty$ at $x$, i.e., elements of 
the presheaf (\ref{0450}) are identified if their restrictions
(namely, the restrictions of their coefficients) to some open
neighbourhood of
$x$ coincide with each other. Let $\cS^*_c$ be the subalgebra of
the BGDA $\cS^*_\infty$ on $X=\Bbb R^n$ which consists of
elements with constant coefficients. Then $\gS^*_x$ is the stalk
of germs of $\cS^*_c$-valued functions on $X$. It is
a bigraded algebra isomorphic to the tensor product
$C^\infty_x\ot_\Bbb R \cS^*_c$ of the
$\Bbb R$-algebra $C^\infty_x$ of the germs of smooth real
functions on $X$ at $x$ and the $\Bbb R$-algebra $\cS^*_c$.
This stalk is naturally decomposed into
$C^\infty_x$-modules $\gS^{k,m}_x$ of the germs of
graded $(k,m)$-forms on $X$. 

Let the common symbol $\Delta$ stand for all the operators
($d$,$d_H$, $d_V$, $\vr$ and $\dl$) on the 
BGDA $\cS^*_\infty$. It is an $\Bbb R$-module morphism whose
restrictions $\Delta_U$ to $\cS^*_U$ intertwined by the
restriction morphisms $r^U_V$ (\ref{0450}) constitute the
direct system of morphisms
\mar{0455}\beq
\{\Delta_U, r^U_V \,\,|\,\, r^U_V\circ\Delta_U=\Delta_V\circ
r^U_V\}, \label{0455}
\eeq
indexed by the directed set of
open neighbourhoods $U$ of $x$. Its direct limit
$\Delta_x$ is an $\Bbb R$-module morphism of the stalk
$\gS^*_x$. The properties of the direct limit of morphisms
are summarized by the following theorem \cite{massey}.

\begin{theo} \label{0460} \mar{0460}
The direct limit of a direct system of complexes $\{C^*_i,i\in
I\}$ is a complex whose cohomology is the direct limit of that
of the complexes $C^*_i$.
\end{theo}

It follows that the stalk
$\gS^*_x$ is a BGDA which contains the complexes corresponding
to the subcomplexes of the variational bicomplex (\ref{0453})
of the BGDA $\cS^*_\infty$.
Since,
$d_x=d_{Hx}+d_{Vx}$ and the operators $d_x$, $d_{Hx}$, $d_{Vx}$
are nilpotent, we have a bicomplex $\gS^{*,*}_x$. Moreover, the
vertical differential
$d_{Vx}$ on
$\gS^*_x=C^\infty_x\ot_\Bbb R \cS^*_c$ comes from the operator
$d_V$ on $\cS^*_c$. Therefore, $\dl_x=\vr_x\circ d_{Vx}$ and
the subcomplexes of $\gS^*_x$ can be assembled into the
variational bicomplex. Accordingly, the sheaf $\gS^*_\infty$ 
 of germs of graded exterior forms on $X$
constitutes the variational bicomplex 
\mar{0452}\beq
(\cO^*_X,\gS^{*,*}_\infty,
\gE_k=\vr(\gS^{k,n}_\infty);d,d_H,d_V,\vr,\dl),
\label{0452}
\eeq
where the operators on
$\gS^*_\infty$ are denoted by the same symbols as those on the
BGDA
$\cS^*_\infty$.

Let $\G(\gS^*_\infty)$ be the bigraded algebra of sections
of the sheaf $\gS^*_\infty$. Given an arbitrary section $s$ of 
$\G(\gS^*_\infty)$, there exists an
open neighbourhood $U$ of each point $x\in X$ such that $s|_U$
is an element of the presheaf (\ref{0450}). It follows that
$\G(\gS^*_\infty)$ can be  provided with the same operators 
$d$, $d_H$, $d_V$, $\vr$ and $\dl$ as the BGDA $\cS^*_\infty$
which make it into the
$\Bbb Z_2$-graded variational bicomplex, analogous to that of
$\cS^*_\infty$. 
The homomorphism $\cS^*_\infty\to\G(\gS^*_\infty)$ is a
monomorphism. 

\bigskip
\noindent
{\it Appendix D}. We quote the
following minor generalization of the abstract de Rham theorem
(\cite{hir}, Theorem 2.12.1) \cite{epr,jmp,tak2}.
Let 
\be
0\to S\ar^h S_0\ar^{h^0} S_1\ar^{h^1}\cdots\ar^{h^{p-1}} S_p\ar^{h^p}
S_{p+1}, \qquad p>1, 
\ee
be an exact sequence of sheaves of abelian groups over a
paracompact topological space $Z$, where the sheaves $S_q$,
$0\leq q<p$, are acyclic, and let 
\beq
0\to \G(Z,S)\ar^{h_*} \G(Z,S_0)\ar^{h^0_*}
\G(Z,S_1)\ar^{h^1_*}\cdots\ar^{h^{p-1}_*} \G(Z,S_p)\ar^{h^p_*}
\G(Z,S_{p+1}) \label{+130}
\eeq
be the corresponding cochain complex
of structure groups of these sheaves.

\begin{theo} \label{+132} \mar{+132}
The $q$-cohomology groups of the
cochain complex (\ref{+130}) for $0\leq q\leq p$ are
isomorphic to the cohomology groups $H^q(Z,S)$ of $Z$ with coefficients in the
sheaf $S$. 
\end{theo}

\end{document}